\documentclass[aps,pre,twocolumn,groupedaddress,showpacs,amsfonts,10pt,%
tightenlines,floatfix]{revtex4-1}

\usepackage[utf8]{inputenc}
\usepackage{graphicx}
\usepackage{amsmath,amssymb}
\usepackage{upgreek}
\usepackage[usenames,dvipsnames]{xcolor}
\usepackage{braket}
\usepackage{cancel}
\usepackage{outlines}

\graphicspath{{./Abbildungen/}}

\begin{document}

\title{A toy model for brain criticality: self-organized excitation/inhibition ratio and the role of network clustering}

\author{Lorenz Baumgarten}
\email[]{lbaumgarten@itp.uni-bremen.de}

\author{Stefan Bornholdt}
\email[]{bornholdt@itp.uni-bremen.de}

\affiliation{Institut für Theoretische Physik, Universität Bremen, 28759 Bremen, Germany}

\date{\today}

\begin{abstract} 
	The critical brain hypothesis receives increasing support from recent experimental results. 
	It postulates that the brain is at a critical point between an ordered and a chaotic regime, 
	sometimes referred to as the "edge of chaos." 
	Another central observation of neuroscience is the principle of excitation-inhibition balance: 
	Certain brain networks exhibit a remarkably constant ratio between excitation and inhibition. 
	When this balance is perturbed, the network shifts away from the critical point, 
	as may for example happen during epileptic seizures. 
	However, it is as of yet unclear what mechanisms balance the neural dynamics towards 
	this excitation-inhibition ratio that ensures critical brain dynamics. 

	Here we introduce a simple yet biologically plausible toy model of a self-organized critical 
	neural network with a self-organizing excitation to inhibition ratio. 

	The model only requires a neuron to have local information of its own recent activity 
	and changes connections between neurons accordingly. We find that the network evolves to a state 
    characterized by avalanche distributions following universal scaling laws typical of criticality, 
    and to a specific excitation to inhibition ratio. 
	The model connects the two questions of brain criticality and of a specific excitation/inhibition 
	balance observed in the brain to a common origin or mechanism. From the perspective of the 
	statistical mechanics of such networks, the model uses the excitation/inhibition ratio as 
	control parameter of a phase transition, which enables criticality at arbitrary high connectivities. 
	We find that network clustering plays a crucial role for this phase transition to occur. 
\end{abstract}


\maketitle

Statistical Mechanics has been a long time companion to neuroscience. 
Decades ago, it played a central role in demonstrating how memory and 
computation can emerge from a network of neurons and thereby laid out the 
foundations of a theory of neural computation. Today, where signs of dynamical 
criticality emerge from neurophysiological data, statistical mechanics can, 
quite similarly, provide elements towards a theory of neural criticality. 
Statistical mechanics has been developed as a toolkit in physics for modeling 
interacting many particle systems by means of maximally reductionist models. 
Magnetic atoms that align to each other are represented as purely binary 
variables (with states up or down). For example, the iconic Ising Model 
\cite{lenz1920note, ising1925beitrag} almost looks like a toy model. 
Nevertheless, such models often make predictions about phase transitions 
of matter that match experimental observations with startling accuracy. 

In the same reductionist approach, networks of neurons can be modeled by representing 
the neuron’s activity as either on ($1$) or off ($0$), dropping most biological detail, 
in order to study dynamical mechanisms and phase transitions on the systemic level 
of neural networks. Such a simplified neural network model is surprisingly similar 
to disordered magnetic glasses, or Ising spin glasses, as has been pointed out by 
Hopfield in his seminal paper \cite{Hopfield1982}. He formulated the idea of associative 
memory: storing memories of patterns as states in the energy landscape of a modified model 
of a magnetic spin glass. This paper initiated a field of statistical mechanics of 
neural networks and the theory of neural computation \cite{amit1989, Hertz1991}.
Its success was based on its central idea that artificial neural networks 
based on symmetric synaptic links, which are nonsense from the biological viewpoint, 
enjoy full access to the tools of equilibrium statistical mechanics and spin glass physics 
\cite{amit1985spin,Kinzel1985,gardner1987maximum,mezard1987spin,kinzel1989statistical,zippelius1993statistical}.
It allowed to calculate the memory capacity of neural networks and to characterize 
the phase transition between order vs.\ chaos---or memory vs.\ forgetting---in great detail. 

Explorations into the more realistic asymmetric neural networks turned out to be more difficult. 
Analytical results were mainly achieved in the sparsely asymmetric limit where asymmetry 
and loops do not fully destroy the energy landscape picture of spin glass physics 
\cite{derrida1987exactly,gardner1986structure,gardner1987zero,gardner1988space,gardner1988optimal,kree1988spin, coolen2001statistical_I, coolen2001statistical_II}.
In addition, numerical studies of random neural networks with fully asymmetric links 
exhibited interesting complex dynamics with an order-chaos phase transition 
\cite{sompolinsky1988chaos,cessac1994occurrence,doyon1994bifurcations,stern2014dynamics}.
A similar class of networks, random automata networks or random Boolean networks, 
originally motivated by the idea that gene regulation networks in living cells determine 
their cell type by means of a dynamical attractor \cite{Kauffman1969, Kauffman1993},
added to this phenomenology. They exhibit a similar order-chaos phase transition and 
their dynamics is characterized by fixed points and periodic attractors and interesting 
properties near criticality 
\cite{derrida1986,derrida1986evolution,derrida1986phase,weisbuch1987phase,socolar2003scaling,aldana2003boolean,kaufman2005scaling, drossel2005number, drossel2005number_b, mihaljev2006scaling, drossel2008random}. 
A popular subset of random Boolean networks, the so-called random threshold networks, 
in fact map onto random neural networks with binary states and links 
\cite{kurten1988correspondence,kurten1988critical,rohlf2002,rohlf2008critical,szejka2008,wang2013effects}. 
The prominent dynamical feature of these networks is the transition between a chaotic 
regime at higher connectivities and a quiescent regime for lower connectivity, 
divided by a critical point in the average connectivity, often at or around an average 
degree $K=2$ for random Boolean networks and for neural networks with zero threshold. 

Statistical mechanics has thus created a fundamental understanding of critical dynamics in networks.
This has been of renewed interest for neuroscience since Beggs and Plenz \cite{BeggsPlenz2003} 
discovered power-law distributed activity avalanche profiles suggesting that 
the brain neural networks may also be poised at a critical point.
Subsequent studies have produced increasingly convincing evidence for brain criticality, 
in the form of more critical neuronal avalanche size and duration distributions 
\cite{Mazzoni2007, Gireesh2008, Pasquale2008, Petermann2009, Tetzlaff2010, Yu2011, Friedman2012, Tagliazucchi2012, Pu2013, Priesemann2014, Scott2014, Bellay2015, Massobrio2015, Timme2016, Yada2017, Yu2017, Ma2018, Ponce2018, Yaghoubi2018, Bowen2019, Fontenele2019, Miller2019} 
with scale-invariant profiles \cite{Friedman2012, Shaukat2016, Timme2016, Ponce2018, Miller2019}.
One popular explanation why brains may be poised at a critical point is that criticality 
has been shown to optimize information processing tasks in certain model systems
\cite{Haldeman2005, Shew2011, Timme2016, Li2017}. 
We would like to point out here an alternative hypothesis, which might be the simplest 
after Occam's razor: The brain must function away from both chaos and quiescence, 
regardless of criticality. With phase transitions, statistical physics provides the 
opportunity to stabilize a system in an intermediate region via tuning to the critical 
point, where criticality itself is not the goal, but only the tool. 

Many numerical models have been developed to describe criticality in neural networks, 
starting with simple critical branching models \cite{Haldeman2005, Williams2014}, integrate-and-fire models 
\cite{Eurich2002, Levina2007a, Pasquale2008, Massobrio2015, Wang2011, Poil2012, DallaPorta2019, Haimovici2013, Rocha2018}, 
or models using other types of neurons \cite{Benayoun2010} that try to 
reproduce the observed critical behavior via finely-tuned or realistic parameters.

There is a host of papers that go beyond replicating the critical behavior to 
presenting algorithms that lead the network to self-organized criticality (SOC), 
often using spike timing dependent plasticity 
\cite{Meisel2009, Rubinov2011, Teixeira2014, khoshkhou2019spike}, 
synaptic depression 
\cite{Levina2007b, Levina2009, Millman2010, Shew2015, Campos2017, Zierenberg2018}, 
Hebbian or anti-Hebbian learning 
\cite{Magnasco2009, Arcangelis2006, pellegrini2007activity}, 
axonal and dendritic outgrowth 
\cite{Abbott2007, Tetzlaff2010, Kossio2018} or combinations of these or other methods 
\cite{Uhlig2013, Stepp2015, brochini2016phase, costa2017self, hernandez2017self, GirardiSchappo2020}---although many of these models still require manually fine-tuned parameters, see \cite{zeraati2021} for a review.
Recent models also combine SOC with learning \cite{Scarpetta2018, Kessenich2018, DelPapa2017}.  

Some of these SOC papers acknowledge the importance of a balance between excitation and 
inhibition in a network for criticality \cite{Magnasco2009, Benayoun2010, Wang2011, Poil2012}, 
as has also been observed experimentally \cite{BeggsPlenz2003}. 
A balance between excitation and inhibition in brains had already been theoretically assumed 
\cite{Shadlen1994, Vreeswijk1996, Amit1997, Shadlen1998, Brunel2000} and experimentally shown 
\cite{Shu2003, Wehr2003, Isaacson2011}, see also \cite{Deneve2016} for a review, outside the context of criticality.
The importance of the balance between excitation and inhibition can also be seen by 
the ratio of excitatory to inhibitory nodes being constant, roughly 4:1, among different 
cortical regions, species, and stages of development \cite{Sahara2012}.\\

Self-organized criticality in adaptive Boolean and neural networks has already been established 
\cite{bornholdt2000topological, rohlf2002, bornholdt2003self, gross2008adaptive, gross2009adaptive} 
before Beggs and Plenz's seminal paper discovering signs of criticality in the brain 
\cite{BeggsPlenz2003}. Models bridging the gap between statistical mechanics 
and neuroscience have subsequently been developed \cite{droste2013}, e.g., by combining 
statistical mechanics SOC models \cite{rohlf2002} with neural network adaptation mechanisms 
such as STDP \cite{Meisel2009}. These models self-organize to a connectivity $K=2$ which is the 
critical value for networks with excitation/inhibition balance 1:1---the common value 
historically used in statistical mechanics models of random networks.  
As criticality in brains is not determined by the average degree (which in any case is 
much larger than $K=2$), but instead by the balance between excitation and inhibition, 
it is an interesting question if these statistical mechanics SOC models also function 
at high connectivities and with excitation/inhibition balance as the control parameter. 
Both ingredients are present in some of the neuroscience models mentioned above; however, 
these contain considerable complexity in order to faithfully model real neural networks.
Here, we want to provide a minimal, yet biologically plausible adaptive neural network model 
with as few parameters as possible that can nonetheless self-organize to a critical point 
at biologically relevant high average degrees and is also hopefully simple enough to 
allow it to be studied analytically using the methods of statistical mechanics.

Our model is a simple Markovian threshold network in which historically criticality has only 
been researched at the critical point at low connectivity $K=2$ \cite{Kuerten1988}. 
Based on our recent observation of the existence of $K$-independent critical points in such 
systems at high connectivities \cite{Neto2017, Baumgarten2019}, but dependent on the ratio 
of excitatory to inhibitory connections, we introduce an algorithm that tunes towards such 
a critical point using only locally available information. We show that this algorithm produces 
high-degree critical networks with specific excitation to inhibition ratios in a wide area of parameter space.
Afterwards, we present an extension of the model which more closely resembles biological networks, 
using a constraint on the maximum number of incoming connections per node, as well as a 
refractory period after firing, and show that this extension also produces criticality.
The property of our algorithm to be independent of implementation details points to a universality 
of the underlying mechanism, which suggests that such an algorithm could be used in a 
variety of networks of all levels of complexity. 

\section*{ Algorithm }
\label{Algorithm-section}

We start with a collection of $N$ neurons, whose states take Boolean values, 
randomly placed in a two-dimensional space with periodic boundary conditions.
The probability of a neuron $i$ being active at time step $t+1$ is given by
\begin{align*}
	P[\sigma_i(t+1)=1] &= \frac 1 {1+\exp\left[-2\beta\left(f_i(t) - 0.5\right)\right]}\\
	\text{with } f_i &= \sum\limits_{j=1}^N c_{ij}(t)\sigma_j(t)\nu(t),
\end{align*}
where $\sigma_i \in \{0,1\}$ is the state of neuron $i$, $f_i(t)$ is the neuron $i$'s incoming signal, $c$ is the adjacency matrix (with $c_{ij} \in \{0,1\}$), and $\nu_j=\pm1$ determines whether a neuron $j$ is excitatory or inhibitory. All neurons are updated synchronously.
Initially, all neurons are unconnected and inactive, i.e., $c_{ij}=0$ and $\sigma_i=0$. The noise introduced via the inverse temperature $\beta$ is therefore necessary for the network to escape this initial inactive state. Unless stated otherwise, we pick $\beta=10$, which accomplishes this within reasonable time and does not affect the network dynamics much otherwise.

The algorithm tuning towards criticality adds and removes connections as follows: 
\begin{outline}
    \1 Every $t_\mathrm r$ time steps randomly select a neuron $i$.
    \1 If the neuron $i$ has been continuously active or continuously inactive during the last $t_\mathrm a$ time steps, it gains an incoming link from another neuron $j\ne i$ that is inhibitory or excitatory, respectively. A neuron $j$ that does not have any outgoing connections can also be chosen, as such a neuron is effectively neither inhibitory nor excitatory yet. \\
    The neuron $j$ is chosen as the nearest eligible neuron without a connection to $i$.
    \1 If the neuron $i$ has been neither exclusively active or inactive during the last $t_\mathrm a$ time steps, i.e., if it changed its state, it instead loses its longest incoming link. 
    \1 If a connection was created originating from a neuron $j$ without any outgoing links, that neuron's identity is then set to inhibitory or excitatory depending on whether neuron $i$ had been continuously active or inactive. 
\end{outline}
Unless stated otherwise, we choose $t_\mathrm r=1$.
As the algorithm starts with a connectionless network, the identities, i.e., excitatory/inhibitory, of all neurons are determined dynamically during rewiring.
This algorithm will initially add excitatory links until the noise creates a stable nonzero activity, similar to real developing networks \cite{Jiang2005}.

If connections are formed and removed randomly, similar to the model we studied in \cite{landmann2021}, instead of creating connections to the nearest eligible neuron and removing the longest connections as described above, the network will be tuned to the well-known critical point at average degree $K=2$.

By keeping connections as short as possible, we can escape this $K=2$ state and tune towards criticality at higher average degrees. This idea is inspired by our finding in \cite{BaumgartenBornholdt2019} that at high degrees a network can be kept in a low-activity state with sensitivity $\lambda\approx1$ if the network has a high clustering coefficient and can also be rationalized biologically by short connections between neurons being more common than long connections.
The sensitivity $\lambda$ is our first indicator for criticality. It is defined as the average number of neurons that will have a different state at time step $t+1$ if a neuron $i$'s state is inverted in time step $t$, i.e., if $i$ is active it is turned inactive and vice versa, than they would otherwise have had \cite{LuqueSole1997, ShmulevichKauffman2004}. If $\lambda>1$, perturbations to the network will on average increase, or they will decrease for $\lambda<1$. The border between these two regimes, $\lambda=1$, is the critical point.

\begin{figure}
	\centering
	\includegraphics[width=1\linewidth]{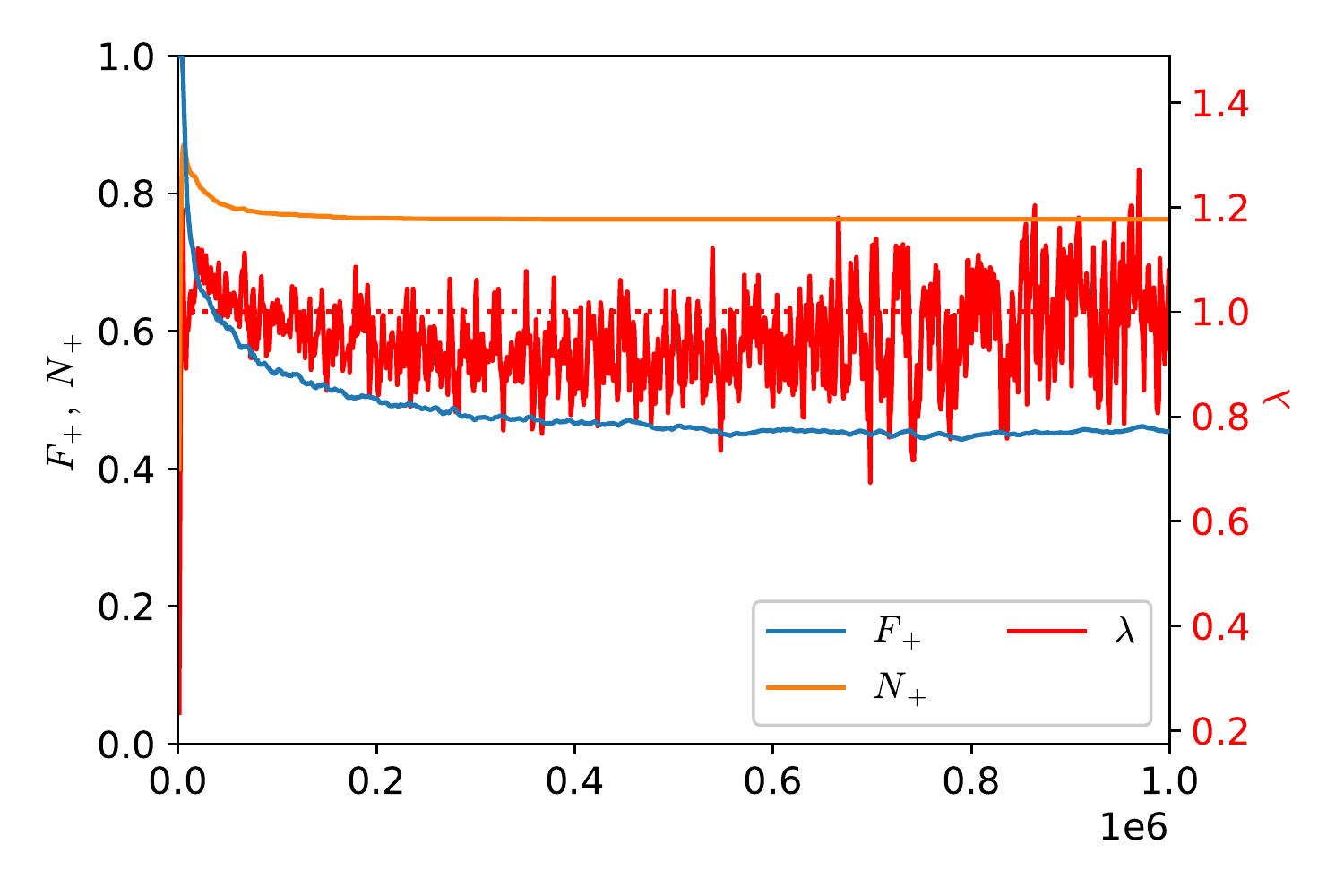}
	\caption{A typical run of the algorithm for $N=2000$ and $t_\mathrm{a}=10^3$. The upper diagram shows the sensitivity $\lambda$ (right axis), the fraction of excitatory connections $F_+$, and the fraction of excitatory nodes $N_+$ (left axis), as a function of $t$. The dotted red line shows the critical sensitivity $\lambda=1$.}%
\label{fig:TuningRun}
\end{figure}

We find that the algorithm maintains a sensitivity $\lambda$ near one while steadily increasing the average degree $K$. A typical run of the algorithm is shown in Figure \ref{fig:TuningRun}.
Increasing the considered time frame $t_\mathrm{a}$ or decreasing the inverse temperature $\beta$ leads to a slower increase in $K$---or no increase at all if nodes are likely to have changed their states within the time window through noise alone---but still produces sensitivities near one if $t_\mathrm{a}$ and $\beta$ are not too large or too small.

\section*{Criticality}
\begin{figure}
	\centering
	\includegraphics[width=1\linewidth]{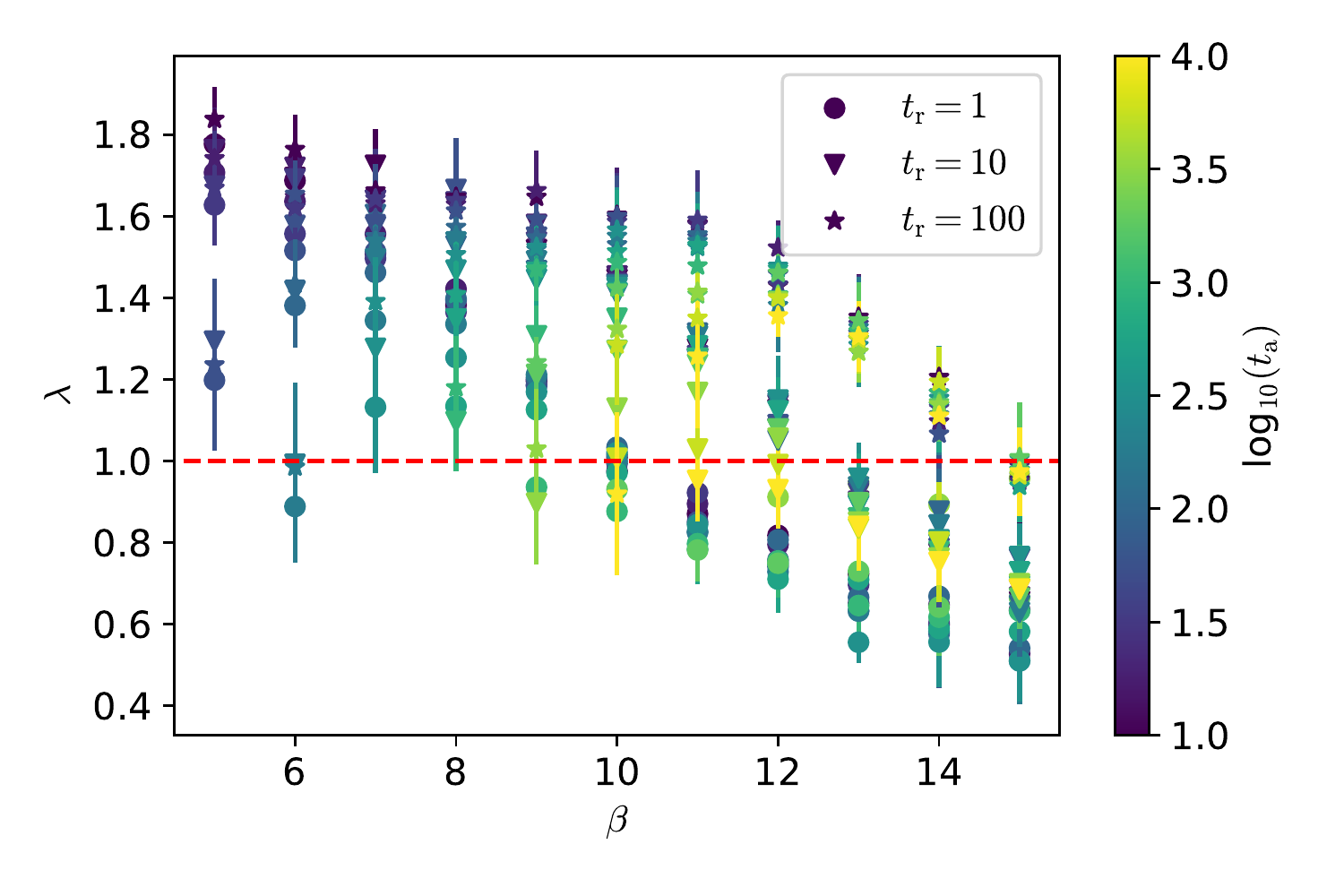}
	\caption{Sensitivity $\lambda$ for different values of the noise $\beta$, the window of states considered for rewiring $t_\mathrm{a}$ and the time between rewiring steps $t_\mathrm{r}$ ($N=2000$, at $K=45$). The red dotted line shows the critical value of $\lambda=1$. Each point is the average of 10 simulations. Parameter combinations in which the network's average degree $K$ did not increase within $10^6\cdot t_\mathrm r$ time steps before reaching $K=45$ in any of the 10 simulations, i.e., if a network with $K=45$ could not reliably be produced within reasonable time, are not shown.}
	\label{fig:Tuning_Scan}
\end{figure}

Figure~\ref{fig:TuningRun} shows that the algorithm does not tune precisely to $\lambda=1$, but since $\lambda=1$ is merely an indicator of criticality and can be inaccurate for clustered networks, this need not discourage us. Additionally, Figure \ref{fig:Tuning_Scan} shows that the sensitivity $\lambda$ does not stray too far from one in a large parameter space.

To further test for criticality, we study activity avalanches in our networks. To measure avalanches, we first let the algorithm run until a target value of $K$ is reached. Here, we pick $K=45\approx\sqrt N$ for $N=2000$. Different choices of $K$ and $N$ do not affect the results much, as long as $K\ll N$ and $t_\mathrm r$ is changed accordingly. The alternative to stopping the algorithm at a fixed $K$ would be to wait for it to arrive at a stationary point, which takes a long time and also only yields average degrees of the order of $N$.
Once the target value of $K$ is reached, we let the network's dynamics continue without any further rewiring and without noise. We then copy the network and flip a node $i$ in the copy at time $t_0$. Next we let both networks run in parallel and measure the time it takes for them to arrive at the same state at a time $t'>t_0$. This is the avalanche duration $T=t'-t_0$. The sum of Hamming distances between the two networks during the avalanche is the avalanche size $S$. The measurement is repeated for every node $i$ in a network and for many different networks.

It is of course possible that the manipulated copy ends up in a different attractor than the original, that it ends up in the same attractor but with a time shift compared to the original, or that the two networks only reach the same state after a large amount of time steps (we stop the measurement after $10^4$ time steps). Since distance and duration are more difficult to define in these cases, we do not use them for our measurements. Our measurements show that this is a relatively rare case -- depending on parameters at most for about 30\% of avalanches.

\begin{figure}[htpb]
	\centering
	\includegraphics[width=1\linewidth]{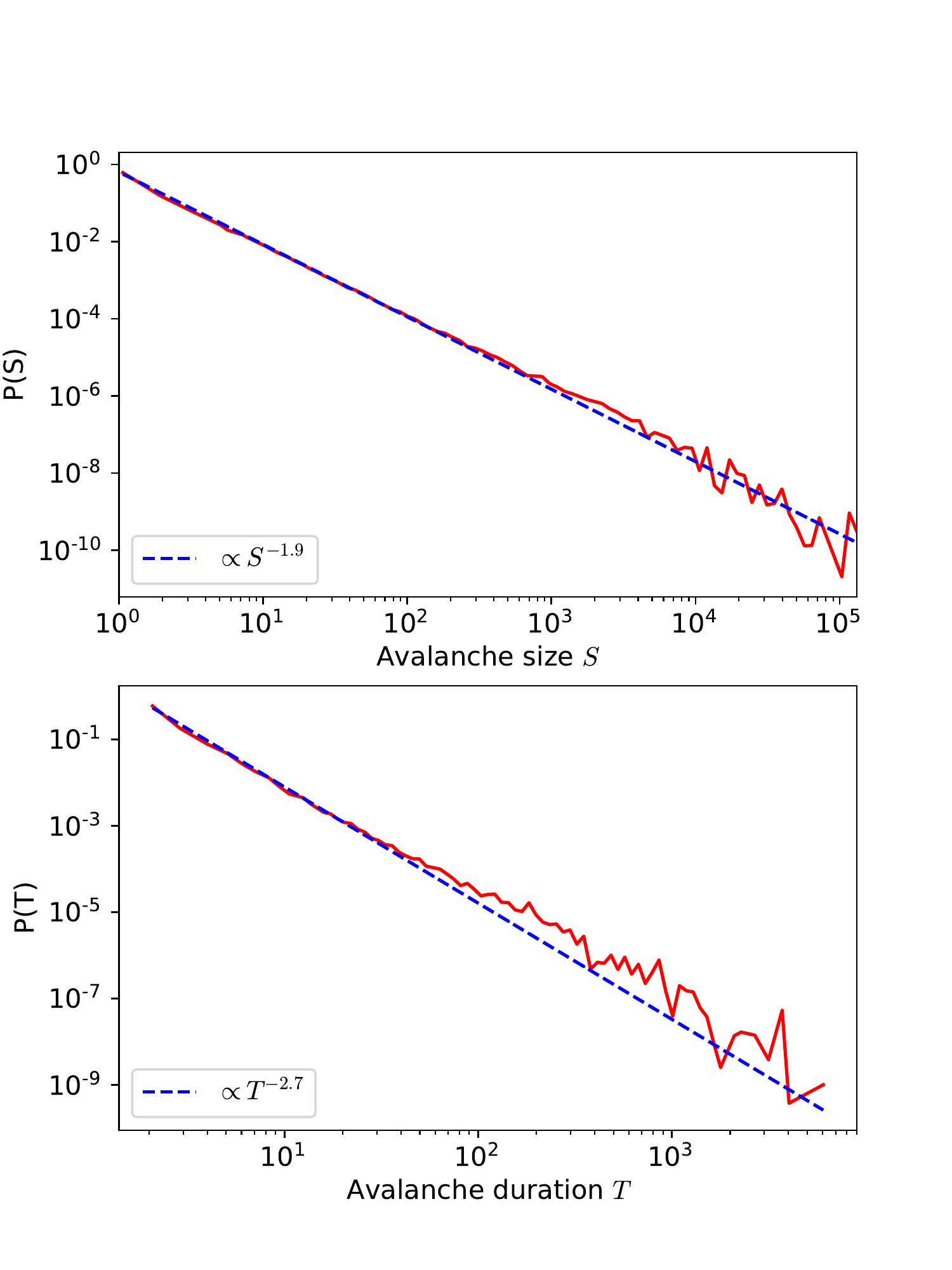}
	\caption{Logarithmically binned avalanche size (upper diagram) and duration (lower diagram) distributions for $K=45$, $N=2000$, and $t_\mathrm{a}=10^3$. The dashed lines show a power-law fit.}%
	\label{fig:Avalanches}
\end{figure}

\begin{figure}
	\includegraphics[width=1\linewidth]{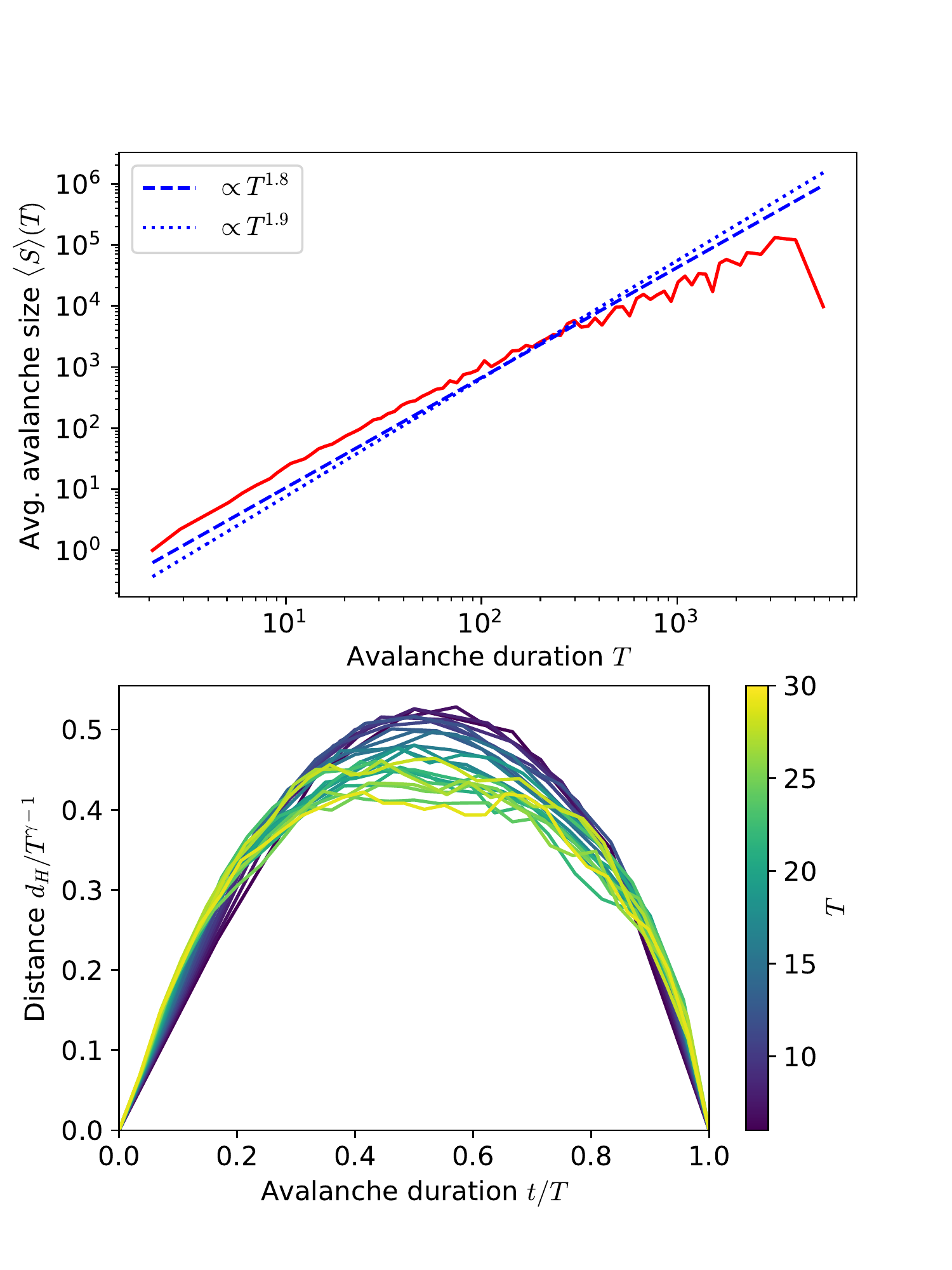}
	\caption{Average avalanche size as a function of avalanche duration (upper diagram) and collapse of avalanche profiles for avalanches of duration 6--30 (lower diagram) for the avalanches shown in Figure \ref{fig:Avalanches}. The dashed line shows a power-law fit, and the dotted line shows a power-law with the theoretical value of $\gamma$ given by equation \eqref{eq:scale_relation}.}%
	\label{fig:Collapse}
\end{figure}

At a critical point, the avalanche size $S$ and avalanche duration $T$ distributions, $P(S)$ and $P(T)$, as well as average avalanche size as a function of avalanche duration should follow power laws \cite{sethna2001}
\begin{align}
	P(S) &\propto S^{-\tau}\\
	P(T) &\propto T^{-\alpha}\\
	\braket S (T) &\propto T^{\gamma},\\
	\text{with } \frac {\alpha-1}{\tau-1}&=\gamma.
	\label{eq:scale_relation}
\end{align}
Further, the avalanche profiles, i.e., Hamming distance to the unperturbed network as a function of time, should collapse onto each other if time is rescaled by the avalanche duration and the Hamming distance $d_\mathrm{H}$ is rescaled by $T^{\gamma-1}$. The three power-laws and the avalanche collapse are shown in Figures~\ref{fig:Avalanches} and \ref{fig:Collapse} for $K=45$. We find approximate power-laws and a sufficient collapse of avalanches, indicating criticality.
The values for $\tau$, $\alpha$, and $\gamma$ indicated by the blue, dashed lines in Figure~\ref{fig:Avalanches} are
\begin{align*}
	\tau&\approx1.8767\pm0.0003\\
    \alpha&\approx2.6916\pm0.0006\\
	\gamma&\approx1.80\pm0.03,\\
	\text{and } \frac {\alpha-1}{\tau-1} &= 1.9296\pm0.0004 \approx \gamma.
\end{align*}
These values have been fitted using the estimator for discrete integer variables described in \cite{Clauset2009}, and the errors given are those resulting from this fitting method.
We have verified that our algorithm also produces approximate power-laws following the scaling relation \eqref{eq:scale_relation} and showing a data collapse for most of the parameter space shown in Figure \ref{fig:Tuning_Scan}. The exponents found vary between $\tau\approx1.6$ and $\tau\approx2.6$ as well as $\alpha\approx2.2$ and $\alpha\approx3.4$.

\begin{figure}
    \centering
    \includegraphics[width=1\linewidth]{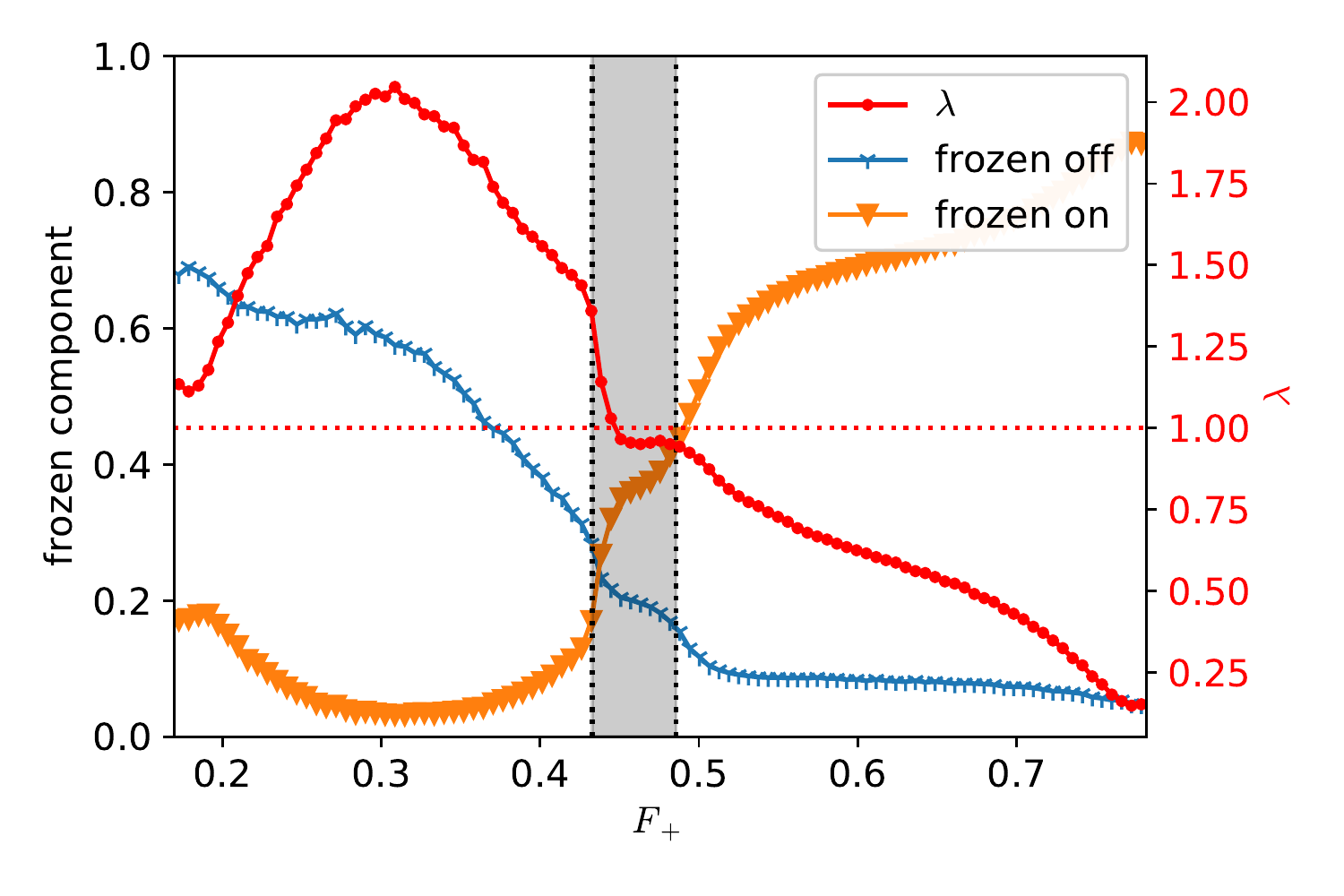}
    \caption{Average frozen on/off component and sensitivity $\lambda$ (right axis) for networks moving from a state perturbed in the $F_+$ space to a critical state ($N=2000$, $K=45$, $t_a=1000$) as a function of the ratio of excitatory links $F_+$. Shown are the averages of 100 networks being perturbed to either side, and the red dotted line shows the critical value of $\lambda$. The frozen on/off component is the fraction of nodes at any point in time that was exclusively on/off during the last $t_a$ time steps. $F_+$ was perturbed by $\pm0.3$ at $K_\mathrm{ini}=25$ and the simulation was ended at $K_\mathrm{fin}=45$. During the simulation, $F_+$ moves from the outer boundaries of the diagram towards the grey area, which shows the area between maximum and minimum values of $F_+$ that networks reached at $K_\mathrm{fin}$, i.e., the critical area of $F_+$ the algorithm tunes towards. The exact value of $F_+$ that is reached is slightly dependent on initial conditions and $F_+$ also fluctuates slightly even after reaching the critical point.}
    \label{fig:Fplus_Run}
\end{figure}

Lastly, we study the critical point's vicinity in the $F_+$ space.
\newpage
To observe the effect of the fraction of excitatory connections $F_+$ on criticality, we use the following procedure:
\begin{outline}
    \1 Evolve a network up to an initial average degree $K_\mathrm{ini}$ using the previously described algorithm
    \1 Perturb the network by either increasing or decreasing $F_+$. In order to do this, repeat the following steps until the desired value of $F_+$ is reached:
    \2 Pick a random neuron
    \2 If $F_+$ shall be increased/decreased and the neuron has an incoming inhibitory/excitatory link:
    \3 Remove the farthest incoming inhibitory/excitatory link
    \3 Form a new incoming excitatory/inhibitory link from the nearest eligible neuron.
    \2 Otherwise, do nothing.
    \1 Resume the previous tuning algorithm until a final average degree $K_\mathrm{fin}$ is reached.
\end{outline}
The algorithm quickly returns the network to a critical state, regardless of $F_+$ at $K_\mathrm{ini}$, and we can observe the sensitivity and frozen components on the way the algorithm takes from the perturbed state to the critical state, as shown in Figure \ref{fig:Fplus_Run}. Note that the perturbation changes $F_+$ but conserves all neurons' in-degrees.

Figure \ref{fig:Fplus_Run} illustrates the functionality of our algorithm. For $F_+$ below the critical point, the frozen off component is larger than the frozen on component, meaning that more nodes are permanently off than on, causing the algorithm to create more excitatory links and thereby increasing $F_+$. The opposite can be seen for $F_+$ above the critical point. The frozen on component is still larger than the frozen off component in the region of $F_+$ the algorithm tunes to. From this, one might expect that $F_+$ would be further decreased here; however, since due to the low total frozen component, many connections are also being removed, if on average more inhibitory connections are being removed than excitatory ones---due to inhibitory connections on average being longer because of the higher out-degree of inhibitory nodes---, $F_+$ must not necessarily decrease within the grey region.\\
We also see that $\lambda$ increases to values significantly above one for $F_+$ below the critical point and significantly below one above the critical point, meaning that it is possible to tune through the critical point, which is another indicator of criticality.\\
Our simulations also show that critical avalanche profiles can still be achieved if the rewiring rule removing connections of flickering nodes is omitted; however, we then lose the ability to tune through the critical point, and the network's ability to return to the critical point after perturbation is diminished because the only way for the network to change its $F_+$ is by adding new links. If we for example constrain the maximum number of links a neuron has, as we will discuss in the following section, returning to a critical state after a perturbation at a saturated degree would therefore be impossible for the network without the ability to remove links.

\section*{Biologically motivated model extensions}
So far we have kept our model as minimal as possible in order to study 
the pure mechanisms of self-organized criticality and excitation/inhibiton regulation. 
We now want to demonstrate that the model can be easily expanded to more closely 
align with properties of realistic neural networks---namely the physiological limits 
on connectivity and the refractory nature of real neurons---without losing its 
ability to self-organize to a critical state. 

Let us start with considering constraints on the resulting average degree $K$. 
As mentioned before, in earlier, non-spatial variations of our model, the algorithm would tune to the critical point $K=2$ \cite{landmann2021}; however, the algorithm presented here does not tune $K$, but instead the ratio of excitation to inhibition via $F_+$ and $N_+$. Therefore, to produce criticality, the connection-removing and connection-producing rules of our algorithm need not balance out, and thus $K$ rises. To keep the average degree from increasing almost indefinitely, we add an additional rule to the model: A node's number of incoming connections cannot exceed a limit $K_\mathrm{max}$.
Such a rule can easily be motivated biologically. Firstly, a brain has reason to be parsimonious with its resources and therefore limit synapses if possible. Secondly, in a biological network, a neuron can simply not have an infinite amount of connections due to spatial restrictions.

The second natural expansion of our model, a refractory period of neuronal activity, has two beneficial side effects. In the base model, the average activity of the network is pushed towards 50\,\%, as connections are added to push nodes away from being permanently active or inactive. This is of course unrealistic for brains, as neurons---as long as they are not part of a spiking avalanche---tend to be inactive apart from occasional spiking due to background noise. This is also reflected in our initial definition of avalanches.
Our avalanches are not avalanches of activity as is common, but instead of distance to an unperturbed comparison network. Both of these points are ameliorated by introducing a refractory period $t_\mathrm{ref}$ to our model as follows:

Nodes cannot be active for $t_\mathrm{ref}$ time steps after being active for one time step. This change is inspired by biology, and nodes can now either be considered as single, primitive spiking neurons or as clusters of neurons which "tire out" and need to recover after spike trains.

We choose a refractory period of $t_\mathrm{ref}=2$ as our observations have shown that $t_\mathrm{ref}=1$ simply produces clusters of two alternatingly blinking parts enabling sustained activity. Any refractory period $t_\mathrm{ref}>1$ does not produce this effect, and therefore we choose the lowest possible value for simplicity's sake. Our algorithm then produces a network whose default state in the absense of activating noise is inactive---although small clusters of sustained activity can still occur, but these are not the norm and often collapse under noise.

The refractory period also requires an adjustment of our rewiring rules as nodes cannot be permanently active anymore: Instead, nodes gain an incoming inhibitory connection if they have ever been active during the last $t_\mathrm a$ time steps and connections are never removed. As previously discussed, this diminishes the model's ability to return to criticality after perturbation, but it is sufficient to show that criticality can be achieved with the model extensions we want to present here. Of course, the rule for removing connections could still be implemented by setting average activity thresholds, but we were unwilling to add more parameters to our model.

\begin{figure}
    \centering
    \includegraphics[width=1\linewidth]{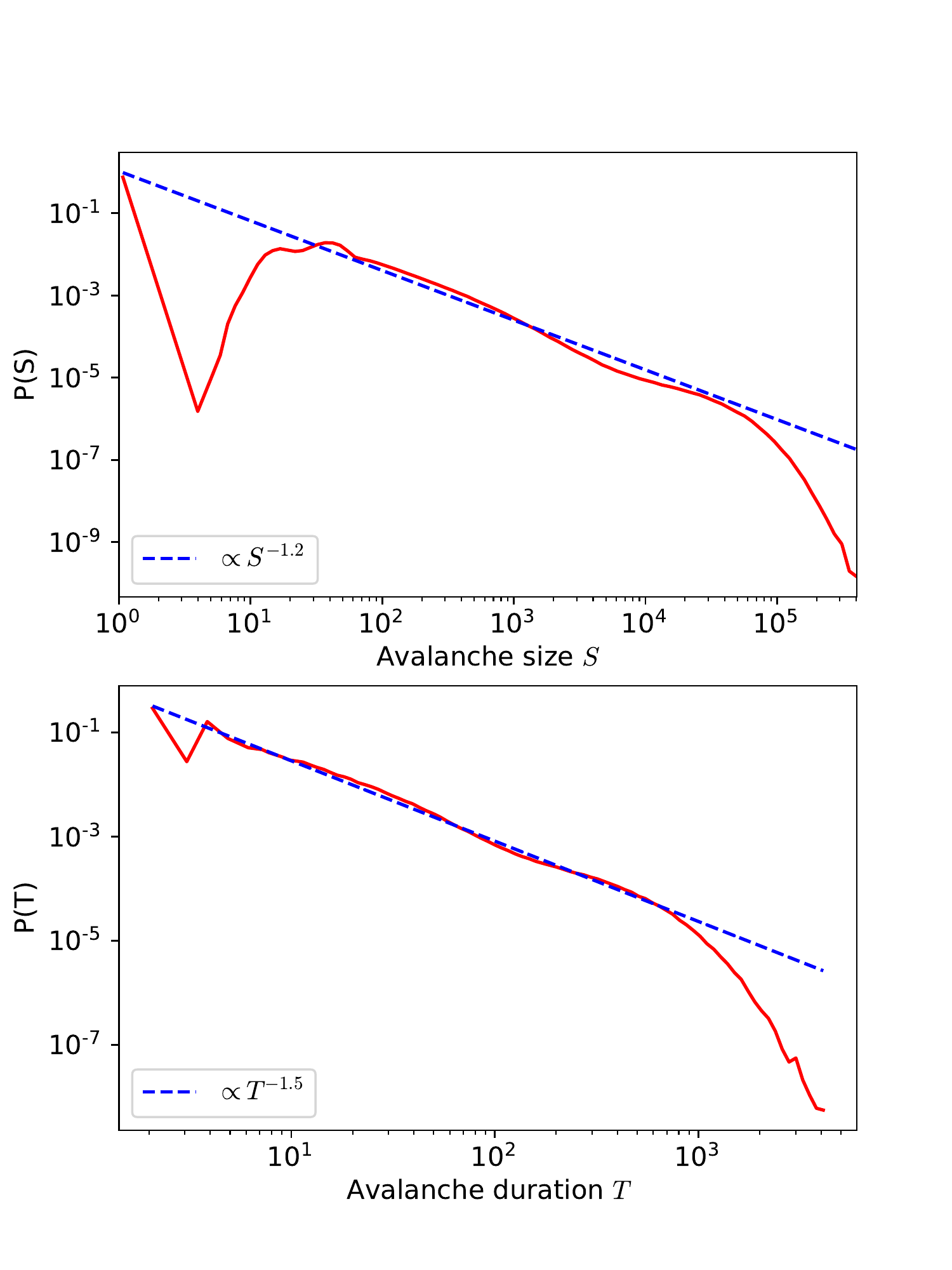}
    \caption{Logarithmically binned avalanche size (upper diagram) and duration (lower diagram) distributions of the model with refractory periods for $K=45$, $N=2000$, $t_\mathrm{a}=10^2$, and $t_\mathrm{r}=10^2$. The dashed lines show a power-law fit.}
    \label{fig:ref_time_Avalanches}
\end{figure}

\begin{figure}
    \centering
    \includegraphics[width=1\linewidth]{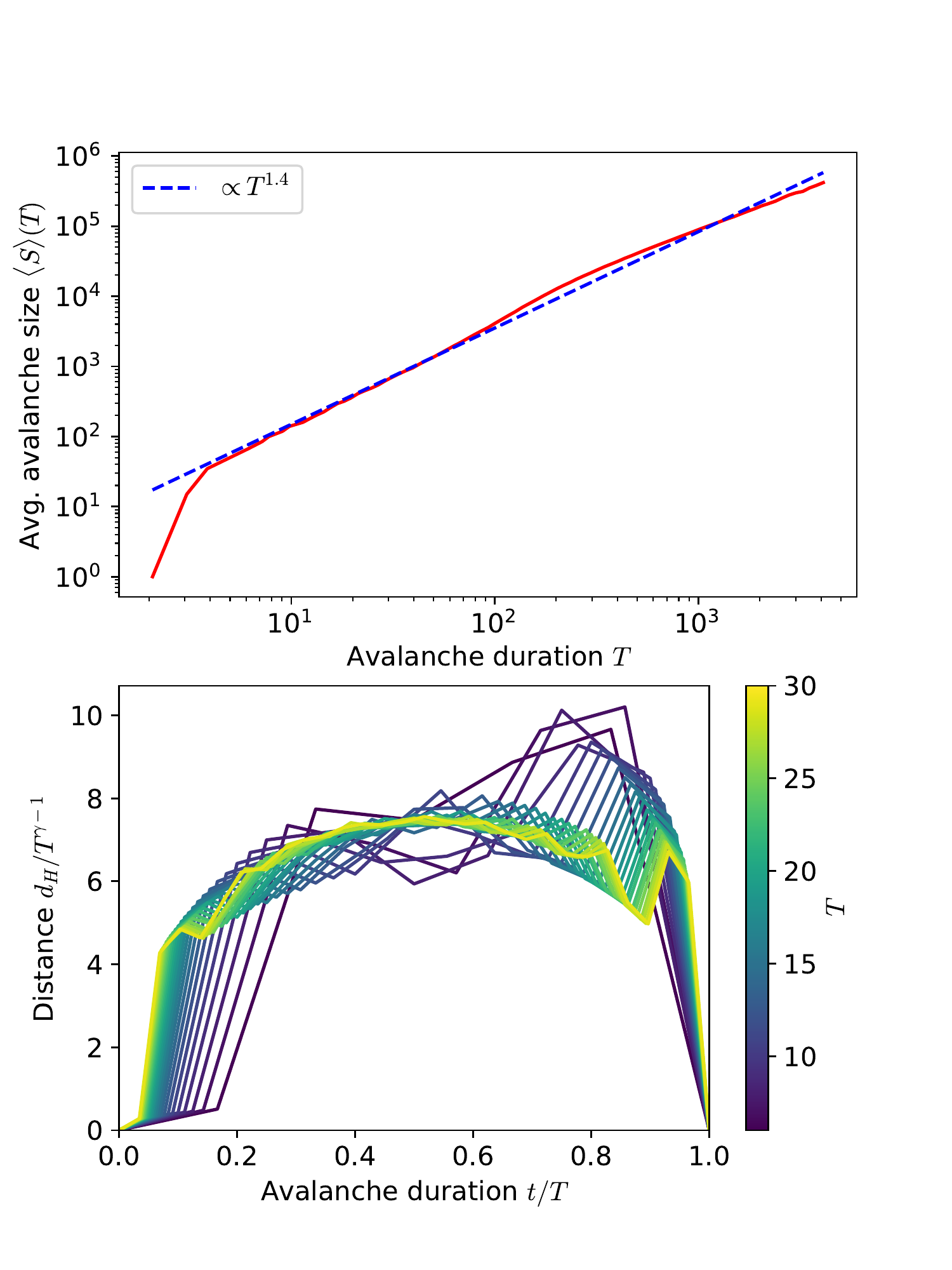}
    \caption{Average avalanche size as a function of avalanche duration (upper diagram) and collapse of avalanche profiles for avalanches of duration 6--30 (lower diagram) for the avalanches shown in Figure \ref{fig:Avalanches}. The dashed line shows a power-law fit.}
    \label{fig:ref_time_Collapse}
\end{figure}

Additionally, the sensitivity is no longer an acceptable indicator of criticality in this case because in the inactive state, the sensitivity is simply the average number of excitatory connections per node which is significantly larger than one. Only once an avalanche starts, does the refractory period prevent nodes from being active consecutively, and it therefore effectively lowers the sensitivity during an avalanche. This again leads to criticality as shown via power-laws and avalanche collapses in Figures \ref{fig:ref_time_Avalanches} and \ref{fig:ref_time_Collapse}. When testing other combinations of $t_\mathrm r$, $t_\mathrm a$, and $\beta$, we found that these parameters still need not be fine-tuned for this model to self-organize to a critical point. The model will reach a critical, high-degree state as long as all of these parameters are sufficiently large.

Since this network operates near a completely inactive state, we can now use a simpler, more intuitive definition of avalanches than before. We start with the network being completely inactive and then activate one node.
The avalanche is then simulated until the network returns to the completely inactive
state and its size is measured as the number of nodes being activated, where a node can also be counted multiple times if it has been active multiple times during the avalanche. Therefore, the avalanches here are true activity avalanches and no longer require a comparison to an unperturbed network.\\
In Figure \ref{fig:ref_time_Collapse}, we can see that the first time step after a node has been activated usually causes a higher activity than subsequent time steps because an excitatory node will, in an inactive network, activate all of its neighbors, but after the first time step, an active node's neighbors may be in their refractory period, thereby reducing the number of nodes that can be activated. This also causes the initial dip in the avalanche size and duration distributions seen in Figure \ref{fig:ref_time_Avalanches}. The last time step before the network returns to inactivity also often shows a high activity. This is because, when a large number of nodes is activated, it is likely that a proportionately large number of inhibiting nodes is activated. Since inhibiting nodes have on average significantly more outgoing connections than excitatory nodes, although more excitatory nodes may be activated, a node in proximity to this activation is likely connected to many of the active inhibitory nodes but only a few of the active excitatory nodes and may therefore not activate. This leads avalanches to often end after a time step with high activity. These high activity steps and subsequent dropoffs---albeit not to complete inactivity---also occur during avalanches but are averaged out in the avalanche profile.

\section*{Conclusion}
In this paper we develop a simple self-organized critical neural network model 
to tune towards the newly discovered critical point at high average degrees discussed in 
\cite{Neto2017, Baumgarten2019}. This critical point, unlike the previously known critical 
point at average degree $K=2$, is nearly independent of the average degree and instead 
depends on the balance between excitation and inhibition. We have thus developed a simple 
algorithm that produces criticality in neural networks at high average degrees, using only 
local information and very few parameters. The algorithm differs from early physics 
papers, which studied the aforementioned critical point at a constant excitation/inhibition ratio 1:1
\cite{derrida1986, Kuerten1988, rohlf2002, szejka2008}, and also from neuroscience models, 
which---due to their closeness to biological reality---are more complex than our model. 

The core idea behind our model is to move the network away from quiescence by adding 
excitatory/inhibitory connections to permanently inactive/active neurons, and to move
the network away from chaos by removing connections from neurons switching their states. 
The addition of two-dimensional space and creating connections to the nearest neurons 
generates network clustering which ultimately allows the model to adjust to the 
critical point with high average degree.

As a result, the model exhibits power-law shaped distributions of activity avalanche sizes 
and durations, which obey universal scaling relationships and can be collapsed onto each other,
the criteria for criticality required by dynamical scaling theory. The model allows to tune 
through the critical point, from a supercritical to a subcritical regime, by varying the ratio 
of excitatory to inhibitory connections in the network. We have also confirmed that the algorithm 
produces the observed scaling for a large part of parameter space. Variants of the model 
that more closely resemble properties of biological networks can be easily built, as we demonstrated 
with an extended version that constrains the maximum number of connections of a node,
and by introducing a refractory period as a simple representation of firing neurons. 
This extension also resulted in networks in a critical state, further indicating that the 
precise implementation of our model is irrelevant for the emergence of criticality.

There already exists a host of neuroscience papers modeling self-organized criticality, 
however, the commonly complex nature and/or high number of parameters, see for example 
\cite{Rubinov2011, Stepp2015, DelPapa2017, GirardiSchappo2020}, of the models and 
the breadth of possible implementations used---such as synaptic depression 
\cite{Levina2007b, Millman2010, Campos2017, khoshkhou2019spike, GirardiSchappo2020}, 
Hebbian or anti-Hebbian learning \cite{Arcangelis2006, pellegrini2007activity, Magnasco2009}, 
STDP \cite{Meisel2009, Teixeira2014, Stepp2015, DelPapa2017}, 
or axonal outgrowth \cite{Abbott2007, Tetzlaff2010, Kossio2018}---makes it difficult 
to gauge which properties of the models are essential for self-organized criticality,  
whereas the model presented here has been trimmed down to its minimal possible version. 
For example, nearly all neuroscience models of self-organized criticality utilize 
integrate-and-fire neurons or other parameter-heavy biologically realistic neuron behaviors 
\cite{Lin2005, Abbott2007, Levina2007b, Levina2009, Meisel2009, Millman2010, Tetzlaff2010, Teixeira2014, Shew2015, Stepp2015, brochini2016phase, costa2017self, hernandez2017self, khoshkhou2019spike}, 
but these, as well as their exact implementation, seem to merely be biological flavor 
not needed for criticality, as shown by our model.
Many of these models opt for less realistic network structures than the one we used, such as fully-connected
\cite{Levina2007b, Magnasco2009, Shew2015, brochini2016phase, costa2017self, khoshkhou2019spike, GirardiSchappo2020} 
or Watts-Strogatz networks \cite{Lin2005, Teixeira2014, Pazzini2021}, and some models omit inhibition 
\cite{Abbott2007, Levina2007b, Campos2017, Pazzini2021} 
or require very specific parameters to attain self-organized criticality 
\cite{Levina2007b, Teixeira2014, Stepp2015, hernandez2017self, khoshkhou2019spike}. 
The previously existing models closest to the one presented here are outgrowth models 
\cite{Abbott2007, Tetzlaff2010, Kossio2018} in which neurons increase or decrease their 
interaction ranges depending on their activity level, but these also require, in addition 
to multiple parameters describing the neuron behavior, manually setting a parameter, 
namely a neuron target activity level, for the network to self-organize to criticality.

In contrast, the minimal nature of our model allows us to isolate and observe the underlying 
mechanism regulating criticality. As a main observation, we find that our model requires 
clustering to achieve the critical state at high connectivities $K$, and that, while a ratio 
of excitatory to inhibitory nodes consolidates during growth, its exact value is not central 
to criticality; rather, criticality is achieved by fine-tuning the connectivity between nodes, 
as has been observed experimentally \cite{sukenik2021}.
Further, the exact algorithm parameters---namely how often rewiring operations are performed, 
governed by $t_\mathrm r$, the time window considered for the rewiring rules, $t_\mathrm a$, 
the noise level $\beta$, and the number of neurons $N$---and the neuron activity implementation, 
i.e., being inactive for a refractory period after activation or not, may vary details of 
the resulting network's behavior, such as the critical exponents, but do not impede criticality 
itself, so long as the parameters are within a rather large area of the parameter space. 
Lastly, to the best of our knowledge, no other self-organized criticality model dynamically 
produces the network's ratio of excitatory to inhibitory nodes.

In addition to pinpointing the mechanisms that enable self-organized criticality,
due to the model's minimal nature, the assumptions we do make about connectivity and clustering 
are realistic for real neural networks within the frame of our modeling approach. 
Since we further showed that the model can be easily extended to include more 
biologically relevant implementation details, such as a refractory period after firing, 
we hope that it can form a useful link between underlying mechanism and more 
detailed models of brain criticality. An interesting question is how the known biological 
and biochemical processes in the brain could implement or interpret the mechanisms studied here. 
We further hope that the simplicity of our model may also encourage analytical follow-up studies 
in statistical mechanics and dynamical systems theory \cite{gross2021not} of self-organized critical neural networks.

\bibliography{references}

\begin{thebibliography}{132}%
\makeatletter
\providecommand \@ifxundefined [1]{%
 \@ifx{#1\undefined}
}%
\providecommand \@ifnum [1]{%
 \ifnum #1\expandafter \@firstoftwo
 \else \expandafter \@secondoftwo
 \fi
}%
\providecommand \@ifx [1]{%
 \ifx #1\expandafter \@firstoftwo
 \else \expandafter \@secondoftwo
 \fi
}%
\providecommand \natexlab [1]{#1}%
\providecommand \enquote  [1]{``#1''}%
\providecommand \bibnamefont  [1]{#1}%
\providecommand \bibfnamefont [1]{#1}%
\providecommand \citenamefont [1]{#1}%
\providecommand \href@noop [0]{\@secondoftwo}%
\providecommand \href [0]{\begingroup \@sanitize@url \@href}%
\providecommand \@href[1]{\@@startlink{#1}\@@href}%
\providecommand \@@href[1]{\endgroup#1\@@endlink}%
\providecommand \@sanitize@url [0]{\catcode `\\12\catcode `\$12\catcode
  `\&12\catcode `\#12\catcode `\^12\catcode `\_12\catcode `\%12\relax}%
\providecommand \@@startlink[1]{}%
\providecommand \@@endlink[0]{}%
\providecommand \url  [0]{\begingroup\@sanitize@url \@url }%
\providecommand \@url [1]{\endgroup\@href {#1}{\urlprefix }}%
\providecommand \urlprefix  [0]{URL }%
\providecommand \Eprint [0]{\href }%
\providecommand \doibase [0]{http://dx.doi.org/}%
\providecommand \selectlanguage [0]{\@gobble}%
\providecommand \bibinfo  [0]{\@secondoftwo}%
\providecommand \bibfield  [0]{\@secondoftwo}%
\providecommand \translation [1]{[#1]}%
\providecommand \BibitemOpen [0]{}%
\providecommand \bibitemStop [0]{}%
\providecommand \bibitemNoStop [0]{.\EOS\space}%
\providecommand \EOS [0]{\spacefactor3000\relax}%
\providecommand \BibitemShut  [1]{\csname bibitem#1\endcsname}%
\let\auto@bib@innerbib\@empty
\bibitem [{\citenamefont {Lenz}(1920)}]{lenz1920note}%
  \BibitemOpen
  \bibfield  {author} {\bibinfo {author} {\bibfnamefont {W.}~\bibnamefont
  {Lenz}},\ }\href@noop {} {\bibfield  {journal} {\bibinfo  {journal} {Z.
  Physik}\ }\textbf {\bibinfo {volume} {21}},\ \bibinfo {pages} {613} (\bibinfo
  {year} {1920})}\BibitemShut {NoStop}%
\bibitem [{\citenamefont {Ising}(1925)}]{ising1925beitrag}%
  \BibitemOpen
  \bibfield  {author} {\bibinfo {author} {\bibfnamefont {E.}~\bibnamefont
  {Ising}},\ }\href@noop {} {\bibfield  {journal} {\bibinfo  {journal} {Z.
  Phys.}\ }\textbf {\bibinfo {volume} {31}},\ \bibinfo {pages} {253} (\bibinfo
  {year} {1925})}\BibitemShut {NoStop}%
\bibitem [{\citenamefont {Hopfield}(1982)}]{Hopfield1982}%
  \BibitemOpen
  \bibfield  {author} {\bibinfo {author} {\bibfnamefont {J.~J.}\ \bibnamefont
  {Hopfield}},\ }\href@noop {} {\bibfield  {journal} {\bibinfo  {journal}
  {Proceedings of the national academy of sciences}\ }\textbf {\bibinfo
  {volume} {79}},\ \bibinfo {pages} {2554} (\bibinfo {year}
  {1982})}\BibitemShut {NoStop}%
\bibitem [{\citenamefont {Amit}(1989)}]{amit1989}%
  \BibitemOpen
  \bibfield  {author} {\bibinfo {author} {\bibfnamefont {D.~J.}\ \bibnamefont
  {Amit}},\ }\href {\doibase 10.1017/CBO9780511623257} {\emph {\bibinfo {title}
  {Modeling Brain Function: The World of Attractor Neural Networks}}}\
  (\bibinfo  {publisher} {Cambridge University Press},\ \bibinfo {year}
  {1989})\BibitemShut {NoStop}%
\bibitem [{\citenamefont {Hertz}\ \emph {et~al.}(1991)\citenamefont {Hertz},
  \citenamefont {Krogh},\ and\ \citenamefont {Palmer}}]{Hertz1991}%
  \BibitemOpen
  \bibfield  {author} {\bibinfo {author} {\bibfnamefont {J.}~\bibnamefont
  {Hertz}}, \bibinfo {author} {\bibfnamefont {A.}~\bibnamefont {Krogh}}, \ and\
  \bibinfo {author} {\bibfnamefont {R.~G.}\ \bibnamefont {Palmer}},\
  }\href@noop {} {\enquote {\bibinfo {title} {Introduction to the theory of
  neural computation},}\ } (\bibinfo {year} {1991})\BibitemShut {NoStop}%
\bibitem [{\citenamefont {Amit}\ \emph {et~al.}(1985)\citenamefont {Amit},
  \citenamefont {Gutfreund},\ and\ \citenamefont {Sompolinsky}}]{amit1985spin}%
  \BibitemOpen
  \bibfield  {author} {\bibinfo {author} {\bibfnamefont {D.~J.}\ \bibnamefont
  {Amit}}, \bibinfo {author} {\bibfnamefont {H.}~\bibnamefont {Gutfreund}}, \
  and\ \bibinfo {author} {\bibfnamefont {H.}~\bibnamefont {Sompolinsky}},\
  }\href@noop {} {\bibfield  {journal} {\bibinfo  {journal} {Physical Review
  A}\ }\textbf {\bibinfo {volume} {32}},\ \bibinfo {pages} {1007} (\bibinfo
  {year} {1985})}\BibitemShut {NoStop}%
\bibitem [{\citenamefont {Kinzel}(1985)}]{Kinzel1985}%
  \BibitemOpen
  \bibfield  {author} {\bibinfo {author} {\bibfnamefont {W.}~\bibnamefont
  {Kinzel}},\ }in\ \href@noop {} {\emph {\bibinfo {booktitle} {Complex Systems
  --- Operational Approaches in Neurobiology, Physics, and Computers}}},\
  \bibinfo {editor} {edited by\ \bibinfo {editor} {\bibfnamefont
  {H.}~\bibnamefont {Haken}}}\ (\bibinfo  {publisher} {Springer Berlin
  Heidelberg},\ \bibinfo {address} {Berlin, Heidelberg},\ \bibinfo {year}
  {1985})\ pp.\ \bibinfo {pages} {107--115}\BibitemShut {NoStop}%
\bibitem [{\citenamefont {Gardner}(1987)}]{gardner1987maximum}%
  \BibitemOpen
  \bibfield  {author} {\bibinfo {author} {\bibfnamefont {E.}~\bibnamefont
  {Gardner}},\ }\href@noop {} {\bibfield  {journal} {\bibinfo  {journal} {EPL
  (Europhysics Letters)}\ }\textbf {\bibinfo {volume} {4}},\ \bibinfo {pages}
  {481} (\bibinfo {year} {1987})}\BibitemShut {NoStop}%
\bibitem [{\citenamefont {M{\'e}zard}\ \emph {et~al.}(1987)\citenamefont
  {M{\'e}zard}, \citenamefont {Parisi},\ and\ \citenamefont
  {Virasoro}}]{mezard1987spin}%
  \BibitemOpen
  \bibfield  {author} {\bibinfo {author} {\bibfnamefont {M.}~\bibnamefont
  {M{\'e}zard}}, \bibinfo {author} {\bibfnamefont {G.}~\bibnamefont {Parisi}},
  \ and\ \bibinfo {author} {\bibfnamefont {M.~A.}\ \bibnamefont {Virasoro}},\
  }\href@noop {} {\emph {\bibinfo {title} {Spin glass theory and beyond: An
  Introduction to the Replica Method and Its Applications}}},\ Vol.~\bibinfo
  {volume} {9}\ (\bibinfo  {publisher} {World Scientific Publishing Company},\
  \bibinfo {year} {1987})\BibitemShut {NoStop}%
\bibitem [{\citenamefont {Kinzel}(1989)}]{kinzel1989statistical}%
  \BibitemOpen
  \bibfield  {author} {\bibinfo {author} {\bibfnamefont {W.}~\bibnamefont
  {Kinzel}},\ }\href@noop {} {\bibfield  {journal} {\bibinfo  {journal}
  {Physica Scripta}\ }\textbf {\bibinfo {volume} {1989}},\ \bibinfo {pages}
  {144} (\bibinfo {year} {1989})}\BibitemShut {NoStop}%
\bibitem [{\citenamefont {Zippelius}(1993)}]{zippelius1993statistical}%
  \BibitemOpen
  \bibfield  {author} {\bibinfo {author} {\bibfnamefont {A.}~\bibnamefont
  {Zippelius}},\ }\href {\doibase https://doi.org/10.1016/0378-4371(93)90378-H}
  {\bibfield  {journal} {\bibinfo  {journal} {Physica A: Statistical Mechanics
  and its Applications}\ }\textbf {\bibinfo {volume} {194}},\ \bibinfo {pages}
  {471} (\bibinfo {year} {1993})}\BibitemShut {NoStop}%
\bibitem [{\citenamefont {Derrida}\ \emph {et~al.}(1987)\citenamefont
  {Derrida}, \citenamefont {Gardner},\ and\ \citenamefont
  {Zippelius}}]{derrida1987exactly}%
  \BibitemOpen
  \bibfield  {author} {\bibinfo {author} {\bibfnamefont {B.}~\bibnamefont
  {Derrida}}, \bibinfo {author} {\bibfnamefont {E.}~\bibnamefont {Gardner}}, \
  and\ \bibinfo {author} {\bibfnamefont {A.}~\bibnamefont {Zippelius}},\
  }\href@noop {} {\bibfield  {journal} {\bibinfo  {journal} {EPL (Europhysics
  Letters)}\ }\textbf {\bibinfo {volume} {4}},\ \bibinfo {pages} {167}
  (\bibinfo {year} {1987})}\BibitemShut {NoStop}%
\bibitem [{\citenamefont {Gardner}(1986)}]{gardner1986structure}%
  \BibitemOpen
  \bibfield  {author} {\bibinfo {author} {\bibfnamefont {E.}~\bibnamefont
  {Gardner}},\ }\href@noop {} {\bibfield  {journal} {\bibinfo  {journal}
  {Journal of Physics A: Mathematical and General}\ }\textbf {\bibinfo {volume}
  {19}},\ \bibinfo {pages} {L1047} (\bibinfo {year} {1986})}\BibitemShut
  {NoStop}%
\bibitem [{\citenamefont {Gardner}\ \emph {et~al.}(1987)\citenamefont
  {Gardner}, \citenamefont {Derrida},\ and\ \citenamefont
  {Mottishaw}}]{gardner1987zero}%
  \BibitemOpen
  \bibfield  {author} {\bibinfo {author} {\bibfnamefont {E.}~\bibnamefont
  {Gardner}}, \bibinfo {author} {\bibfnamefont {B.}~\bibnamefont {Derrida}}, \
  and\ \bibinfo {author} {\bibfnamefont {P.}~\bibnamefont {Mottishaw}},\
  }\href@noop {} {\bibfield  {journal} {\bibinfo  {journal} {Journal de
  physique}\ }\textbf {\bibinfo {volume} {48}},\ \bibinfo {pages} {741}
  (\bibinfo {year} {1987})}\BibitemShut {NoStop}%
\bibitem [{\citenamefont {Gardner}(1988)}]{gardner1988space}%
  \BibitemOpen
  \bibfield  {author} {\bibinfo {author} {\bibfnamefont {E.}~\bibnamefont
  {Gardner}},\ }\href@noop {} {\bibfield  {journal} {\bibinfo  {journal}
  {Journal of physics A: Mathematical and general}\ }\textbf {\bibinfo {volume}
  {21}},\ \bibinfo {pages} {257} (\bibinfo {year} {1988})}\BibitemShut
  {NoStop}%
\bibitem [{\citenamefont {Gardner}\ and\ \citenamefont
  {Derrida}(1988)}]{gardner1988optimal}%
  \BibitemOpen
  \bibfield  {author} {\bibinfo {author} {\bibfnamefont {E.}~\bibnamefont
  {Gardner}}\ and\ \bibinfo {author} {\bibfnamefont {B.}~\bibnamefont
  {Derrida}},\ }\href@noop {} {\bibfield  {journal} {\bibinfo  {journal}
  {Journal of Physics A: Mathematical and general}\ }\textbf {\bibinfo {volume}
  {21}},\ \bibinfo {pages} {271} (\bibinfo {year} {1988})}\BibitemShut
  {NoStop}%
\bibitem [{\citenamefont {Kree}\ \emph {et~al.}(1988)\citenamefont {Kree},
  \citenamefont {Widmaier},\ and\ \citenamefont {Zippelius}}]{kree1988spin}%
  \BibitemOpen
  \bibfield  {author} {\bibinfo {author} {\bibfnamefont {R.}~\bibnamefont
  {Kree}}, \bibinfo {author} {\bibfnamefont {D.}~\bibnamefont {Widmaier}}, \
  and\ \bibinfo {author} {\bibfnamefont {A.}~\bibnamefont {Zippelius}},\
  }\href@noop {} {\bibfield  {journal} {\bibinfo  {journal} {Journal of Physics
  A: Mathematical and General}\ }\textbf {\bibinfo {volume} {21}},\ \bibinfo
  {pages} {L1181} (\bibinfo {year} {1988})}\BibitemShut {NoStop}%
\bibitem [{\citenamefont
  {Coolen}(2001{\natexlab{a}})}]{coolen2001statistical_I}%
  \BibitemOpen
  \bibfield  {author} {\bibinfo {author} {\bibfnamefont {A.}~\bibnamefont
  {Coolen}},\ }in\ \href@noop {} {\emph {\bibinfo {booktitle} {Handbook of
  biological physics}}},\ Vol.~\bibinfo {volume} {4}\ (\bibinfo  {publisher}
  {Elsevier},\ \bibinfo {year} {2001})\ pp.\ \bibinfo {pages}
  {553--618}\BibitemShut {NoStop}%
\bibitem [{\citenamefont
  {Coolen}(2001{\natexlab{b}})}]{coolen2001statistical_II}%
  \BibitemOpen
  \bibfield  {author} {\bibinfo {author} {\bibfnamefont {A.}~\bibnamefont
  {Coolen}},\ }in\ \href@noop {} {\emph {\bibinfo {booktitle} {Handbook of
  biological physics}}},\ Vol.~\bibinfo {volume} {4}\ (\bibinfo  {publisher}
  {Elsevier},\ \bibinfo {year} {2001})\ pp.\ \bibinfo {pages}
  {619--684}\BibitemShut {NoStop}%
\bibitem [{\citenamefont {Sompolinsky}\ \emph {et~al.}(1988)\citenamefont
  {Sompolinsky}, \citenamefont {Crisanti},\ and\ \citenamefont
  {Sommers}}]{sompolinsky1988chaos}%
  \BibitemOpen
  \bibfield  {author} {\bibinfo {author} {\bibfnamefont {H.}~\bibnamefont
  {Sompolinsky}}, \bibinfo {author} {\bibfnamefont {A.}~\bibnamefont
  {Crisanti}}, \ and\ \bibinfo {author} {\bibfnamefont {H.-J.}\ \bibnamefont
  {Sommers}},\ }\href@noop {} {\bibfield  {journal} {\bibinfo  {journal}
  {Physical review letters}\ }\textbf {\bibinfo {volume} {61}},\ \bibinfo
  {pages} {259} (\bibinfo {year} {1988})}\BibitemShut {NoStop}%
\bibitem [{\citenamefont {Cessac}(1994)}]{cessac1994occurrence}%
  \BibitemOpen
  \bibfield  {author} {\bibinfo {author} {\bibfnamefont {B.}~\bibnamefont
  {Cessac}},\ }\href@noop {} {\bibfield  {journal} {\bibinfo  {journal} {EPL
  (Europhysics Letters)}\ }\textbf {\bibinfo {volume} {26}},\ \bibinfo {pages}
  {577} (\bibinfo {year} {1994})}\BibitemShut {NoStop}%
\bibitem [{\citenamefont {Doyon}\ \emph {et~al.}(1994)\citenamefont {Doyon},
  \citenamefont {Cessac}, \citenamefont {Quoy},\ and\ \citenamefont
  {Samuelides}}]{doyon1994bifurcations}%
  \BibitemOpen
  \bibfield  {author} {\bibinfo {author} {\bibfnamefont {B.}~\bibnamefont
  {Doyon}}, \bibinfo {author} {\bibfnamefont {B.}~\bibnamefont {Cessac}},
  \bibinfo {author} {\bibfnamefont {M.}~\bibnamefont {Quoy}}, \ and\ \bibinfo
  {author} {\bibfnamefont {M.}~\bibnamefont {Samuelides}},\ }\href@noop {}
  {\bibfield  {journal} {\bibinfo  {journal} {Acta biotheoretica}\ }\textbf
  {\bibinfo {volume} {42}},\ \bibinfo {pages} {215} (\bibinfo {year}
  {1994})}\BibitemShut {NoStop}%
\bibitem [{\citenamefont {Stern}\ \emph {et~al.}(2014)\citenamefont {Stern},
  \citenamefont {Sompolinsky},\ and\ \citenamefont
  {Abbott}}]{stern2014dynamics}%
  \BibitemOpen
  \bibfield  {author} {\bibinfo {author} {\bibfnamefont {M.}~\bibnamefont
  {Stern}}, \bibinfo {author} {\bibfnamefont {H.}~\bibnamefont {Sompolinsky}},
  \ and\ \bibinfo {author} {\bibfnamefont {L.}~\bibnamefont {Abbott}},\
  }\href@noop {} {\bibfield  {journal} {\bibinfo  {journal} {Physical Review
  E}\ }\textbf {\bibinfo {volume} {90}},\ \bibinfo {pages} {062710} (\bibinfo
  {year} {2014})}\BibitemShut {NoStop}%
\bibitem [{\citenamefont {Kauffman}(1969)}]{Kauffman1969}%
  \BibitemOpen
  \bibfield  {author} {\bibinfo {author} {\bibfnamefont {S.~A.}\ \bibnamefont
  {Kauffman}},\ }\href@noop {} {\bibfield  {journal} {\bibinfo  {journal}
  {Journal of theoretical biology}\ }\textbf {\bibinfo {volume} {22}},\
  \bibinfo {pages} {437} (\bibinfo {year} {1969})}\BibitemShut {NoStop}%
\bibitem [{\citenamefont {Kauffman}\ \emph {et~al.}(1993)\citenamefont
  {Kauffman} \emph {et~al.}}]{Kauffman1993}%
  \BibitemOpen
  \bibfield  {author} {\bibinfo {author} {\bibfnamefont {S.~A.}\ \bibnamefont
  {Kauffman}} \emph {et~al.},\ }\href@noop {} {\emph {\bibinfo {title} {The
  origins of order: Self-organization and selection in evolution}}}\ (\bibinfo
  {publisher} {Oxford University Press, USA},\ \bibinfo {year}
  {1993})\BibitemShut {NoStop}%
\bibitem [{\citenamefont {Derrida}\ and\ \citenamefont
  {Pomeau}(1986)}]{derrida1986}%
  \BibitemOpen
  \bibfield  {author} {\bibinfo {author} {\bibfnamefont {B.}~\bibnamefont
  {Derrida}}\ and\ \bibinfo {author} {\bibfnamefont {Y.}~\bibnamefont
  {Pomeau}},\ }\href@noop {} {\bibfield  {journal} {\bibinfo  {journal} {EPL
  (Europhysics Letters)}\ }\textbf {\bibinfo {volume} {1}},\ \bibinfo {pages}
  {45} (\bibinfo {year} {1986})}\BibitemShut {NoStop}%
\bibitem [{\citenamefont {Derrida}\ and\ \citenamefont
  {Weisbuch}(1986)}]{derrida1986evolution}%
  \BibitemOpen
  \bibfield  {author} {\bibinfo {author} {\bibfnamefont {B.}~\bibnamefont
  {Derrida}}\ and\ \bibinfo {author} {\bibfnamefont {G.}~\bibnamefont
  {Weisbuch}},\ }\href@noop {} {\bibfield  {journal} {\bibinfo  {journal}
  {Journal de physique}\ }\textbf {\bibinfo {volume} {47}},\ \bibinfo {pages}
  {1297} (\bibinfo {year} {1986})}\BibitemShut {NoStop}%
\bibitem [{\citenamefont {Derrida}\ and\ \citenamefont
  {Stauffer}(1986)}]{derrida1986phase}%
  \BibitemOpen
  \bibfield  {author} {\bibinfo {author} {\bibfnamefont {B.}~\bibnamefont
  {Derrida}}\ and\ \bibinfo {author} {\bibfnamefont {D.}~\bibnamefont
  {Stauffer}},\ }\href@noop {} {\bibfield  {journal} {\bibinfo  {journal} {EPL
  (Europhysics Letters)}\ }\textbf {\bibinfo {volume} {2}},\ \bibinfo {pages}
  {739} (\bibinfo {year} {1986})}\BibitemShut {NoStop}%
\bibitem [{\citenamefont {Weisbuch}\ and\ \citenamefont
  {Stauffer}(1987)}]{weisbuch1987phase}%
  \BibitemOpen
  \bibfield  {author} {\bibinfo {author} {\bibfnamefont {G.}~\bibnamefont
  {Weisbuch}}\ and\ \bibinfo {author} {\bibfnamefont {D.}~\bibnamefont
  {Stauffer}},\ }\href@noop {} {\bibfield  {journal} {\bibinfo  {journal}
  {Journal de physique}\ }\textbf {\bibinfo {volume} {48}},\ \bibinfo {pages}
  {11} (\bibinfo {year} {1987})}\BibitemShut {NoStop}%
\bibitem [{\citenamefont {Socolar}\ and\ \citenamefont
  {Kauffman}(2003)}]{socolar2003scaling}%
  \BibitemOpen
  \bibfield  {author} {\bibinfo {author} {\bibfnamefont {J.~E.~S.}\
  \bibnamefont {Socolar}}\ and\ \bibinfo {author} {\bibfnamefont {S.~A.}\
  \bibnamefont {Kauffman}},\ }\href {\doibase 10.1103/PhysRevLett.90.068702}
  {\bibfield  {journal} {\bibinfo  {journal} {Phys. Rev. Lett.}\ }\textbf
  {\bibinfo {volume} {90}},\ \bibinfo {pages} {068702} (\bibinfo {year}
  {2003})}\BibitemShut {NoStop}%
\bibitem [{\citenamefont {Aldana}\ \emph {et~al.}(2003)\citenamefont {Aldana},
  \citenamefont {Coppersmith},\ and\ \citenamefont
  {Kadanoff}}]{aldana2003boolean}%
  \BibitemOpen
  \bibfield  {author} {\bibinfo {author} {\bibfnamefont {M.}~\bibnamefont
  {Aldana}}, \bibinfo {author} {\bibfnamefont {S.}~\bibnamefont {Coppersmith}},
  \ and\ \bibinfo {author} {\bibfnamefont {L.~P.}\ \bibnamefont {Kadanoff}},\
  }\href@noop {} {\bibfield  {journal} {\bibinfo  {journal} {Perspectives and
  Problems in Nolinear Science}\ ,\ \bibinfo {pages} {23}} (\bibinfo {year}
  {2003})}\BibitemShut {NoStop}%
\bibitem [{\citenamefont {Kaufman}\ \emph {et~al.}(2005)\citenamefont
  {Kaufman}, \citenamefont {Mihaljev},\ and\ \citenamefont
  {Drossel}}]{kaufman2005scaling}%
  \BibitemOpen
  \bibfield  {author} {\bibinfo {author} {\bibfnamefont {V.}~\bibnamefont
  {Kaufman}}, \bibinfo {author} {\bibfnamefont {T.}~\bibnamefont {Mihaljev}}, \
  and\ \bibinfo {author} {\bibfnamefont {B.}~\bibnamefont {Drossel}},\
  }\href@noop {} {\bibfield  {journal} {\bibinfo  {journal} {Physical Review
  E}\ }\textbf {\bibinfo {volume} {72}},\ \bibinfo {pages} {046124} (\bibinfo
  {year} {2005})}\BibitemShut {NoStop}%
\bibitem [{\citenamefont {Drossel}(2005)}]{drossel2005number}%
  \BibitemOpen
  \bibfield  {author} {\bibinfo {author} {\bibfnamefont {B.}~\bibnamefont
  {Drossel}},\ }\href@noop {} {\bibfield  {journal} {\bibinfo  {journal}
  {Physical Review E}\ }\textbf {\bibinfo {volume} {72}},\ \bibinfo {pages}
  {016110} (\bibinfo {year} {2005})}\BibitemShut {NoStop}%
\bibitem [{\citenamefont {Drossel}\ \emph {et~al.}(2005)\citenamefont
  {Drossel}, \citenamefont {Mihaljev},\ and\ \citenamefont
  {Greil}}]{drossel2005number_b}%
  \BibitemOpen
  \bibfield  {author} {\bibinfo {author} {\bibfnamefont {B.}~\bibnamefont
  {Drossel}}, \bibinfo {author} {\bibfnamefont {T.}~\bibnamefont {Mihaljev}}, \
  and\ \bibinfo {author} {\bibfnamefont {F.}~\bibnamefont {Greil}},\
  }\href@noop {} {\bibfield  {journal} {\bibinfo  {journal} {Physical review
  letters}\ }\textbf {\bibinfo {volume} {94}},\ \bibinfo {pages} {088701}
  (\bibinfo {year} {2005})}\BibitemShut {NoStop}%
\bibitem [{\citenamefont {Mihaljev}\ and\ \citenamefont
  {Drossel}(2006)}]{mihaljev2006scaling}%
  \BibitemOpen
  \bibfield  {author} {\bibinfo {author} {\bibfnamefont {T.}~\bibnamefont
  {Mihaljev}}\ and\ \bibinfo {author} {\bibfnamefont {B.}~\bibnamefont
  {Drossel}},\ }\href@noop {} {\bibfield  {journal} {\bibinfo  {journal}
  {Physical Review E}\ }\textbf {\bibinfo {volume} {74}},\ \bibinfo {pages}
  {046101} (\bibinfo {year} {2006})}\BibitemShut {NoStop}%
\bibitem [{\citenamefont {Drossel}(2008)}]{drossel2008random}%
  \BibitemOpen
  \bibfield  {author} {\bibinfo {author} {\bibfnamefont {B.}~\bibnamefont
  {Drossel}},\ }\href@noop {} {\bibfield  {journal} {\bibinfo  {journal}
  {Reviews of nonlinear dynamics and complexity 1}\ ,\ \bibinfo {pages} {69}}
  (\bibinfo {year} {2008})}\BibitemShut {NoStop}%
\bibitem [{\citenamefont {Kurten}(1988)}]{kurten1988correspondence}%
  \BibitemOpen
  \bibfield  {author} {\bibinfo {author} {\bibfnamefont {K.}~\bibnamefont
  {Kurten}},\ }\href@noop {} {\bibfield  {journal} {\bibinfo  {journal}
  {Journal of Physics A: Mathematical and General}\ }\textbf {\bibinfo {volume}
  {21}},\ \bibinfo {pages} {L615} (\bibinfo {year} {1988})}\BibitemShut
  {NoStop}%
\bibitem [{\citenamefont
  {K{\"u}rten}(1988{\natexlab{a}})}]{kurten1988critical}%
  \BibitemOpen
  \bibfield  {author} {\bibinfo {author} {\bibfnamefont {K.~E.}\ \bibnamefont
  {K{\"u}rten}},\ }\href@noop {} {\bibfield  {journal} {\bibinfo  {journal}
  {Physics Letters A}\ }\textbf {\bibinfo {volume} {129}},\ \bibinfo {pages}
  {157} (\bibinfo {year} {1988}{\natexlab{a}})}\BibitemShut {NoStop}%
\bibitem [{\citenamefont {Rohlf}\ and\ \citenamefont
  {Bornholdt}(2002)}]{rohlf2002}%
  \BibitemOpen
  \bibfield  {author} {\bibinfo {author} {\bibfnamefont {T.}~\bibnamefont
  {Rohlf}}\ and\ \bibinfo {author} {\bibfnamefont {S.}~\bibnamefont
  {Bornholdt}},\ }\href@noop {} {\bibfield  {journal} {\bibinfo  {journal}
  {Physica A: Statistical Mechanics and its Applications}\ }\textbf {\bibinfo
  {volume} {310}},\ \bibinfo {pages} {245} (\bibinfo {year}
  {2002})}\BibitemShut {NoStop}%
\bibitem [{\citenamefont {Rohlf}(2008)}]{rohlf2008critical}%
  \BibitemOpen
  \bibfield  {author} {\bibinfo {author} {\bibfnamefont {T.}~\bibnamefont
  {Rohlf}},\ }\href@noop {} {\bibfield  {journal} {\bibinfo  {journal}
  {Physical Review E}\ }\textbf {\bibinfo {volume} {78}},\ \bibinfo {pages}
  {066118} (\bibinfo {year} {2008})}\BibitemShut {NoStop}%
\bibitem [{\citenamefont {Szejka}\ \emph {et~al.}(2008)\citenamefont {Szejka},
  \citenamefont {Mihaljev},\ and\ \citenamefont {Drossel}}]{szejka2008}%
  \BibitemOpen
  \bibfield  {author} {\bibinfo {author} {\bibfnamefont {A.}~\bibnamefont
  {Szejka}}, \bibinfo {author} {\bibfnamefont {T.}~\bibnamefont {Mihaljev}}, \
  and\ \bibinfo {author} {\bibfnamefont {B.}~\bibnamefont {Drossel}},\
  }\href@noop {} {\bibfield  {journal} {\bibinfo  {journal} {New Journal of
  Physics}\ }\textbf {\bibinfo {volume} {10}},\ \bibinfo {pages} {063009}
  (\bibinfo {year} {2008})}\BibitemShut {NoStop}%
\bibitem [{\citenamefont {Wang}\ and\ \citenamefont
  {Albert}(2013)}]{wang2013effects}%
  \BibitemOpen
  \bibfield  {author} {\bibinfo {author} {\bibfnamefont {R.-S.}\ \bibnamefont
  {Wang}}\ and\ \bibinfo {author} {\bibfnamefont {R.}~\bibnamefont {Albert}},\
  }\href@noop {} {\bibfield  {journal} {\bibinfo  {journal} {Physical Review
  E}\ }\textbf {\bibinfo {volume} {87}},\ \bibinfo {pages} {012810} (\bibinfo
  {year} {2013})}\BibitemShut {NoStop}%
\bibitem [{\citenamefont {Beggs}\ and\ \citenamefont
  {Plenz}(2003)}]{BeggsPlenz2003}%
  \BibitemOpen
  \bibfield  {author} {\bibinfo {author} {\bibfnamefont {J.~M.}\ \bibnamefont
  {Beggs}}\ and\ \bibinfo {author} {\bibfnamefont {D.}~\bibnamefont {Plenz}},\
  }\href {\doibase 10.1523/JNEUROSCI.23-35-11167.2003} {\bibfield  {journal}
  {\bibinfo  {journal} {Journal of Neuroscience}\ }\textbf {\bibinfo {volume}
  {23}},\ \bibinfo {pages} {11167} (\bibinfo {year} {2003})},\ \Eprint
  {http://arxiv.org/abs/https://www.jneurosci.org/content/23/35/11167.full.pdf}
  {https://www.jneurosci.org/content/23/35/11167.full.pdf} \BibitemShut
  {NoStop}%
\bibitem [{\citenamefont {Mazzoni}\ \emph {et~al.}(2007)\citenamefont
  {Mazzoni}, \citenamefont {Broccard}, \citenamefont {Garcia-Perez},
  \citenamefont {Bonifazi}, \citenamefont {Ruaro},\ and\ \citenamefont
  {Torre}}]{Mazzoni2007}%
  \BibitemOpen
  \bibfield  {author} {\bibinfo {author} {\bibfnamefont {A.}~\bibnamefont
  {Mazzoni}}, \bibinfo {author} {\bibfnamefont {F.~D.}\ \bibnamefont
  {Broccard}}, \bibinfo {author} {\bibfnamefont {E.}~\bibnamefont
  {Garcia-Perez}}, \bibinfo {author} {\bibfnamefont {P.}~\bibnamefont
  {Bonifazi}}, \bibinfo {author} {\bibfnamefont {M.~E.}\ \bibnamefont {Ruaro}},
  \ and\ \bibinfo {author} {\bibfnamefont {V.}~\bibnamefont {Torre}},\
  }\href@noop {} {\bibfield  {journal} {\bibinfo  {journal} {PloS one}\
  }\textbf {\bibinfo {volume} {2}} (\bibinfo {year} {2007})}\BibitemShut
  {NoStop}%
\bibitem [{\citenamefont {Gireesh}\ and\ \citenamefont
  {Plenz}(2008)}]{Gireesh2008}%
  \BibitemOpen
  \bibfield  {author} {\bibinfo {author} {\bibfnamefont {E.~D.}\ \bibnamefont
  {Gireesh}}\ and\ \bibinfo {author} {\bibfnamefont {D.}~\bibnamefont
  {Plenz}},\ }\href {\doibase 10.1073/pnas.0800537105} {\bibfield  {journal}
  {\bibinfo  {journal} {Proceedings of the National Academy of Sciences}\
  }\textbf {\bibinfo {volume} {105}},\ \bibinfo {pages} {7576} (\bibinfo {year}
  {2008})},\ \Eprint
  {http://arxiv.org/abs/https://www.pnas.org/content/105/21/7576.full.pdf}
  {https://www.pnas.org/content/105/21/7576.full.pdf} \BibitemShut {NoStop}%
\bibitem [{\citenamefont {Pasquale}\ \emph {et~al.}(2008)\citenamefont
  {Pasquale}, \citenamefont {Massobrio}, \citenamefont {Bologna}, \citenamefont
  {Chiappalone},\ and\ \citenamefont {Martinoia}}]{Pasquale2008}%
  \BibitemOpen
  \bibfield  {author} {\bibinfo {author} {\bibfnamefont {V.}~\bibnamefont
  {Pasquale}}, \bibinfo {author} {\bibfnamefont {P.}~\bibnamefont {Massobrio}},
  \bibinfo {author} {\bibfnamefont {L.}~\bibnamefont {Bologna}}, \bibinfo
  {author} {\bibfnamefont {M.}~\bibnamefont {Chiappalone}}, \ and\ \bibinfo
  {author} {\bibfnamefont {S.}~\bibnamefont {Martinoia}},\ }\href@noop {}
  {\bibfield  {journal} {\bibinfo  {journal} {Neuroscience}\ }\textbf {\bibinfo
  {volume} {153}},\ \bibinfo {pages} {1354} (\bibinfo {year}
  {2008})}\BibitemShut {NoStop}%
\bibitem [{\citenamefont {Petermann}\ \emph {et~al.}(2009)\citenamefont
  {Petermann}, \citenamefont {Thiagarajan}, \citenamefont {Lebedev},
  \citenamefont {Nicolelis}, \citenamefont {Chialvo},\ and\ \citenamefont
  {Plenz}}]{Petermann2009}%
  \BibitemOpen
  \bibfield  {author} {\bibinfo {author} {\bibfnamefont {T.}~\bibnamefont
  {Petermann}}, \bibinfo {author} {\bibfnamefont {T.~C.}\ \bibnamefont
  {Thiagarajan}}, \bibinfo {author} {\bibfnamefont {M.~A.}\ \bibnamefont
  {Lebedev}}, \bibinfo {author} {\bibfnamefont {M.~A.}\ \bibnamefont
  {Nicolelis}}, \bibinfo {author} {\bibfnamefont {D.~R.}\ \bibnamefont
  {Chialvo}}, \ and\ \bibinfo {author} {\bibfnamefont {D.}~\bibnamefont
  {Plenz}},\ }\href@noop {} {\bibfield  {journal} {\bibinfo  {journal}
  {Proceedings of the National Academy of Sciences}\ }\textbf {\bibinfo
  {volume} {106}},\ \bibinfo {pages} {15921} (\bibinfo {year}
  {2009})}\BibitemShut {NoStop}%
\bibitem [{\citenamefont {Tetzlaff}\ \emph {et~al.}(2010)\citenamefont
  {Tetzlaff}, \citenamefont {Okujeni}, \citenamefont {Egert}, \citenamefont
  {W{\"o}rg{\"o}tter},\ and\ \citenamefont {Butz}}]{Tetzlaff2010}%
  \BibitemOpen
  \bibfield  {author} {\bibinfo {author} {\bibfnamefont {C.}~\bibnamefont
  {Tetzlaff}}, \bibinfo {author} {\bibfnamefont {S.}~\bibnamefont {Okujeni}},
  \bibinfo {author} {\bibfnamefont {U.}~\bibnamefont {Egert}}, \bibinfo
  {author} {\bibfnamefont {F.}~\bibnamefont {W{\"o}rg{\"o}tter}}, \ and\
  \bibinfo {author} {\bibfnamefont {M.}~\bibnamefont {Butz}},\ }\href@noop {}
  {\bibfield  {journal} {\bibinfo  {journal} {PLoS computational biology}\
  }\textbf {\bibinfo {volume} {6}} (\bibinfo {year} {2010})}\BibitemShut
  {NoStop}%
\bibitem [{\citenamefont {Yu}\ \emph {et~al.}(2011)\citenamefont {Yu},
  \citenamefont {Yang}, \citenamefont {Nakahara}, \citenamefont {Santos},
  \citenamefont {Nikoli{\'c}},\ and\ \citenamefont {Plenz}}]{Yu2011}%
  \BibitemOpen
  \bibfield  {author} {\bibinfo {author} {\bibfnamefont {S.}~\bibnamefont
  {Yu}}, \bibinfo {author} {\bibfnamefont {H.}~\bibnamefont {Yang}}, \bibinfo
  {author} {\bibfnamefont {H.}~\bibnamefont {Nakahara}}, \bibinfo {author}
  {\bibfnamefont {G.~S.}\ \bibnamefont {Santos}}, \bibinfo {author}
  {\bibfnamefont {D.}~\bibnamefont {Nikoli{\'c}}}, \ and\ \bibinfo {author}
  {\bibfnamefont {D.}~\bibnamefont {Plenz}},\ }\href@noop {} {\bibfield
  {journal} {\bibinfo  {journal} {Journal of neuroscience}\ }\textbf {\bibinfo
  {volume} {31}},\ \bibinfo {pages} {17514} (\bibinfo {year}
  {2011})}\BibitemShut {NoStop}%
\bibitem [{\citenamefont {Friedman}\ \emph {et~al.}(2012)\citenamefont
  {Friedman}, \citenamefont {Ito}, \citenamefont {Brinkman}, \citenamefont
  {Shimono}, \citenamefont {DeVille}, \citenamefont {Dahmen}, \citenamefont
  {Beggs},\ and\ \citenamefont {Butler}}]{Friedman2012}%
  \BibitemOpen
  \bibfield  {author} {\bibinfo {author} {\bibfnamefont {N.}~\bibnamefont
  {Friedman}}, \bibinfo {author} {\bibfnamefont {S.}~\bibnamefont {Ito}},
  \bibinfo {author} {\bibfnamefont {B.~A.}\ \bibnamefont {Brinkman}}, \bibinfo
  {author} {\bibfnamefont {M.}~\bibnamefont {Shimono}}, \bibinfo {author}
  {\bibfnamefont {R.~L.}\ \bibnamefont {DeVille}}, \bibinfo {author}
  {\bibfnamefont {K.~A.}\ \bibnamefont {Dahmen}}, \bibinfo {author}
  {\bibfnamefont {J.~M.}\ \bibnamefont {Beggs}}, \ and\ \bibinfo {author}
  {\bibfnamefont {T.~C.}\ \bibnamefont {Butler}},\ }\href@noop {} {\bibfield
  {journal} {\bibinfo  {journal} {Physical review letters}\ }\textbf {\bibinfo
  {volume} {108}},\ \bibinfo {pages} {208102} (\bibinfo {year}
  {2012})}\BibitemShut {NoStop}%
\bibitem [{\citenamefont {Tagliazucchi}\ \emph {et~al.}(2012)\citenamefont
  {Tagliazucchi}, \citenamefont {Balenzuela}, \citenamefont {Fraiman},\ and\
  \citenamefont {Chialvo}}]{Tagliazucchi2012}%
  \BibitemOpen
  \bibfield  {author} {\bibinfo {author} {\bibfnamefont {E.}~\bibnamefont
  {Tagliazucchi}}, \bibinfo {author} {\bibfnamefont {P.}~\bibnamefont
  {Balenzuela}}, \bibinfo {author} {\bibfnamefont {D.}~\bibnamefont {Fraiman}},
  \ and\ \bibinfo {author} {\bibfnamefont {D.~R.}\ \bibnamefont {Chialvo}},\
  }\href@noop {} {\bibfield  {journal} {\bibinfo  {journal} {Frontiers in
  physiology}\ }\textbf {\bibinfo {volume} {3}},\ \bibinfo {pages} {15}
  (\bibinfo {year} {2012})}\BibitemShut {NoStop}%
\bibitem [{\citenamefont {Pu}\ \emph {et~al.}(2013)\citenamefont {Pu},
  \citenamefont {Gong}, \citenamefont {Li},\ and\ \citenamefont
  {Luo}}]{Pu2013}%
  \BibitemOpen
  \bibfield  {author} {\bibinfo {author} {\bibfnamefont {J.}~\bibnamefont
  {Pu}}, \bibinfo {author} {\bibfnamefont {H.}~\bibnamefont {Gong}}, \bibinfo
  {author} {\bibfnamefont {X.}~\bibnamefont {Li}}, \ and\ \bibinfo {author}
  {\bibfnamefont {Q.}~\bibnamefont {Luo}},\ }\href@noop {} {\bibfield
  {journal} {\bibinfo  {journal} {Scientific reports}\ }\textbf {\bibinfo
  {volume} {3}},\ \bibinfo {pages} {1} (\bibinfo {year} {2013})}\BibitemShut
  {NoStop}%
\bibitem [{\citenamefont {Priesemann}\ \emph {et~al.}(2014)\citenamefont
  {Priesemann}, \citenamefont {Wibral}, \citenamefont {Valderrama},
  \citenamefont {Pr{\"o}pper}, \citenamefont {Le~Van~Quyen}, \citenamefont
  {Geisel}, \citenamefont {Triesch}, \citenamefont {Nikoli{\'c}},\ and\
  \citenamefont {Munk}}]{Priesemann2014}%
  \BibitemOpen
  \bibfield  {author} {\bibinfo {author} {\bibfnamefont {V.}~\bibnamefont
  {Priesemann}}, \bibinfo {author} {\bibfnamefont {M.}~\bibnamefont {Wibral}},
  \bibinfo {author} {\bibfnamefont {M.}~\bibnamefont {Valderrama}}, \bibinfo
  {author} {\bibfnamefont {R.}~\bibnamefont {Pr{\"o}pper}}, \bibinfo {author}
  {\bibfnamefont {M.}~\bibnamefont {Le~Van~Quyen}}, \bibinfo {author}
  {\bibfnamefont {T.}~\bibnamefont {Geisel}}, \bibinfo {author} {\bibfnamefont
  {J.}~\bibnamefont {Triesch}}, \bibinfo {author} {\bibfnamefont
  {D.}~\bibnamefont {Nikoli{\'c}}}, \ and\ \bibinfo {author} {\bibfnamefont
  {M.~H.}\ \bibnamefont {Munk}},\ }\href@noop {} {\bibfield  {journal}
  {\bibinfo  {journal} {Frontiers in systems neuroscience}\ }\textbf {\bibinfo
  {volume} {8}},\ \bibinfo {pages} {108} (\bibinfo {year} {2014})}\BibitemShut
  {NoStop}%
\bibitem [{\citenamefont {Scott}\ \emph {et~al.}(2014)\citenamefont {Scott},
  \citenamefont {Fagerholm}, \citenamefont {Mutoh}, \citenamefont {Leech},
  \citenamefont {Sharp}, \citenamefont {Shew},\ and\ \citenamefont
  {Kn{\"o}pfel}}]{Scott2014}%
  \BibitemOpen
  \bibfield  {author} {\bibinfo {author} {\bibfnamefont {G.}~\bibnamefont
  {Scott}}, \bibinfo {author} {\bibfnamefont {E.~D.}\ \bibnamefont
  {Fagerholm}}, \bibinfo {author} {\bibfnamefont {H.}~\bibnamefont {Mutoh}},
  \bibinfo {author} {\bibfnamefont {R.}~\bibnamefont {Leech}}, \bibinfo
  {author} {\bibfnamefont {D.~J.}\ \bibnamefont {Sharp}}, \bibinfo {author}
  {\bibfnamefont {W.~L.}\ \bibnamefont {Shew}}, \ and\ \bibinfo {author}
  {\bibfnamefont {T.}~\bibnamefont {Kn{\"o}pfel}},\ }\href@noop {} {\bibfield
  {journal} {\bibinfo  {journal} {Journal of Neuroscience}\ }\textbf {\bibinfo
  {volume} {34}},\ \bibinfo {pages} {16611} (\bibinfo {year}
  {2014})}\BibitemShut {NoStop}%
\bibitem [{\citenamefont {Bellay}\ \emph {et~al.}(2015)\citenamefont {Bellay},
  \citenamefont {Klaus}, \citenamefont {Seshadri},\ and\ \citenamefont
  {Plenz}}]{Bellay2015}%
  \BibitemOpen
  \bibfield  {author} {\bibinfo {author} {\bibfnamefont {T.}~\bibnamefont
  {Bellay}}, \bibinfo {author} {\bibfnamefont {A.}~\bibnamefont {Klaus}},
  \bibinfo {author} {\bibfnamefont {S.}~\bibnamefont {Seshadri}}, \ and\
  \bibinfo {author} {\bibfnamefont {D.}~\bibnamefont {Plenz}},\ }\href@noop {}
  {\bibfield  {journal} {\bibinfo  {journal} {Elife}\ }\textbf {\bibinfo
  {volume} {4}},\ \bibinfo {pages} {e07224} (\bibinfo {year}
  {2015})}\BibitemShut {NoStop}%
\bibitem [{\citenamefont {Massobrio}\ \emph {et~al.}(2015)\citenamefont
  {Massobrio}, \citenamefont {Pasquale},\ and\ \citenamefont
  {Martinoia}}]{Massobrio2015}%
  \BibitemOpen
  \bibfield  {author} {\bibinfo {author} {\bibfnamefont {P.}~\bibnamefont
  {Massobrio}}, \bibinfo {author} {\bibfnamefont {V.}~\bibnamefont {Pasquale}},
  \ and\ \bibinfo {author} {\bibfnamefont {S.}~\bibnamefont {Martinoia}},\
  }\href@noop {} {\bibfield  {journal} {\bibinfo  {journal} {Scientific
  reports}\ }\textbf {\bibinfo {volume} {5}},\ \bibinfo {pages} {10578}
  (\bibinfo {year} {2015})}\BibitemShut {NoStop}%
\bibitem [{\citenamefont {Timme}\ \emph {et~al.}(2016)\citenamefont {Timme},
  \citenamefont {Marshall}, \citenamefont {Bennett}, \citenamefont {Ripp},
  \citenamefont {Lautzenhiser},\ and\ \citenamefont {Beggs}}]{Timme2016}%
  \BibitemOpen
  \bibfield  {author} {\bibinfo {author} {\bibfnamefont {N.~M.}\ \bibnamefont
  {Timme}}, \bibinfo {author} {\bibfnamefont {N.~J.}\ \bibnamefont {Marshall}},
  \bibinfo {author} {\bibfnamefont {N.}~\bibnamefont {Bennett}}, \bibinfo
  {author} {\bibfnamefont {M.}~\bibnamefont {Ripp}}, \bibinfo {author}
  {\bibfnamefont {E.}~\bibnamefont {Lautzenhiser}}, \ and\ \bibinfo {author}
  {\bibfnamefont {J.~M.}\ \bibnamefont {Beggs}},\ }\href@noop {} {\bibfield
  {journal} {\bibinfo  {journal} {Frontiers in physiology}\ }\textbf {\bibinfo
  {volume} {7}},\ \bibinfo {pages} {425} (\bibinfo {year} {2016})}\BibitemShut
  {NoStop}%
\bibitem [{\citenamefont {Yada}\ \emph {et~al.}(2017)\citenamefont {Yada},
  \citenamefont {Mita}, \citenamefont {Sanada}, \citenamefont {Yano},
  \citenamefont {Kanzaki}, \citenamefont {Bakkum}, \citenamefont {Hierlemann},\
  and\ \citenamefont {Takahashi}}]{Yada2017}%
  \BibitemOpen
  \bibfield  {author} {\bibinfo {author} {\bibfnamefont {Y.}~\bibnamefont
  {Yada}}, \bibinfo {author} {\bibfnamefont {T.}~\bibnamefont {Mita}}, \bibinfo
  {author} {\bibfnamefont {A.}~\bibnamefont {Sanada}}, \bibinfo {author}
  {\bibfnamefont {R.}~\bibnamefont {Yano}}, \bibinfo {author} {\bibfnamefont
  {R.}~\bibnamefont {Kanzaki}}, \bibinfo {author} {\bibfnamefont {D.~J.}\
  \bibnamefont {Bakkum}}, \bibinfo {author} {\bibfnamefont {A.}~\bibnamefont
  {Hierlemann}}, \ and\ \bibinfo {author} {\bibfnamefont {H.}~\bibnamefont
  {Takahashi}},\ }\href@noop {} {\bibfield  {journal} {\bibinfo  {journal}
  {Neuroscience}\ }\textbf {\bibinfo {volume} {343}},\ \bibinfo {pages} {55}
  (\bibinfo {year} {2017})}\BibitemShut {NoStop}%
\bibitem [{\citenamefont {Yu}\ \emph {et~al.}(2017)\citenamefont {Yu},
  \citenamefont {Ribeiro}, \citenamefont {Meisel}, \citenamefont {Chou},
  \citenamefont {Mitz}, \citenamefont {Saunders},\ and\ \citenamefont
  {Plenz}}]{Yu2017}%
  \BibitemOpen
  \bibfield  {author} {\bibinfo {author} {\bibfnamefont {S.}~\bibnamefont
  {Yu}}, \bibinfo {author} {\bibfnamefont {T.~L.}\ \bibnamefont {Ribeiro}},
  \bibinfo {author} {\bibfnamefont {C.}~\bibnamefont {Meisel}}, \bibinfo
  {author} {\bibfnamefont {S.}~\bibnamefont {Chou}}, \bibinfo {author}
  {\bibfnamefont {A.}~\bibnamefont {Mitz}}, \bibinfo {author} {\bibfnamefont
  {R.}~\bibnamefont {Saunders}}, \ and\ \bibinfo {author} {\bibfnamefont
  {D.}~\bibnamefont {Plenz}},\ }\href@noop {} {\bibfield  {journal} {\bibinfo
  {journal} {Elife}\ }\textbf {\bibinfo {volume} {6}},\ \bibinfo {pages}
  {e27119} (\bibinfo {year} {2017})}\BibitemShut {NoStop}%
\bibitem [{\citenamefont {Ma}\ \emph {et~al.}(2018)\citenamefont {Ma},
  \citenamefont {Turrigiano}, \citenamefont {Wessel},\ and\ \citenamefont
  {Hengen}}]{Ma2018}%
  \BibitemOpen
  \bibfield  {author} {\bibinfo {author} {\bibfnamefont {Z.}~\bibnamefont
  {Ma}}, \bibinfo {author} {\bibfnamefont {G.~G.}\ \bibnamefont {Turrigiano}},
  \bibinfo {author} {\bibfnamefont {R.}~\bibnamefont {Wessel}}, \ and\ \bibinfo
  {author} {\bibfnamefont {K.~B.}\ \bibnamefont {Hengen}},\ }\href@noop {}
  {\bibfield  {journal} {\bibinfo  {journal} {bioRxiv}\ ,\ \bibinfo {pages}
  {503243}} (\bibinfo {year} {2018})}\BibitemShut {NoStop}%
\bibitem [{\citenamefont {Ponce-Alvarez}\ \emph {et~al.}(2018)\citenamefont
  {Ponce-Alvarez}, \citenamefont {Jouary}, \citenamefont {Privat},
  \citenamefont {Deco},\ and\ \citenamefont {Sumbre}}]{Ponce2018}%
  \BibitemOpen
  \bibfield  {author} {\bibinfo {author} {\bibfnamefont {A.}~\bibnamefont
  {Ponce-Alvarez}}, \bibinfo {author} {\bibfnamefont {A.}~\bibnamefont
  {Jouary}}, \bibinfo {author} {\bibfnamefont {M.}~\bibnamefont {Privat}},
  \bibinfo {author} {\bibfnamefont {G.}~\bibnamefont {Deco}}, \ and\ \bibinfo
  {author} {\bibfnamefont {G.}~\bibnamefont {Sumbre}},\ }\href@noop {}
  {\bibfield  {journal} {\bibinfo  {journal} {Neuron}\ }\textbf {\bibinfo
  {volume} {100}},\ \bibinfo {pages} {1446} (\bibinfo {year}
  {2018})}\BibitemShut {NoStop}%
\bibitem [{\citenamefont {Yaghoubi}\ \emph {et~al.}(2018)\citenamefont
  {Yaghoubi}, \citenamefont {de~Graaf}, \citenamefont {Orlandi}, \citenamefont
  {Girotto}, \citenamefont {Colicos},\ and\ \citenamefont
  {Davidsen}}]{Yaghoubi2018}%
  \BibitemOpen
  \bibfield  {author} {\bibinfo {author} {\bibfnamefont {M.}~\bibnamefont
  {Yaghoubi}}, \bibinfo {author} {\bibfnamefont {T.}~\bibnamefont {de~Graaf}},
  \bibinfo {author} {\bibfnamefont {J.~G.}\ \bibnamefont {Orlandi}}, \bibinfo
  {author} {\bibfnamefont {F.}~\bibnamefont {Girotto}}, \bibinfo {author}
  {\bibfnamefont {M.~A.}\ \bibnamefont {Colicos}}, \ and\ \bibinfo {author}
  {\bibfnamefont {J.}~\bibnamefont {Davidsen}},\ }\href@noop {} {\bibfield
  {journal} {\bibinfo  {journal} {Scientific reports}\ }\textbf {\bibinfo
  {volume} {8}},\ \bibinfo {pages} {1} (\bibinfo {year} {2018})}\BibitemShut
  {NoStop}%
\bibitem [{\citenamefont {Bowen}\ \emph {et~al.}(2019)\citenamefont {Bowen},
  \citenamefont {Winkowski}, \citenamefont {Seshadri}, \citenamefont {Plenz},\
  and\ \citenamefont {Kanold}}]{Bowen2019}%
  \BibitemOpen
  \bibfield  {author} {\bibinfo {author} {\bibfnamefont {Z.}~\bibnamefont
  {Bowen}}, \bibinfo {author} {\bibfnamefont {D.}~\bibnamefont {Winkowski}},
  \bibinfo {author} {\bibfnamefont {S.}~\bibnamefont {Seshadri}}, \bibinfo
  {author} {\bibfnamefont {D.}~\bibnamefont {Plenz}}, \ and\ \bibinfo {author}
  {\bibfnamefont {P.~O.}\ \bibnamefont {Kanold}},\ }\href@noop {} {\bibfield
  {journal} {\bibinfo  {journal} {Frontiers in systems neuroscience}\ }\textbf
  {\bibinfo {volume} {13}},\ \bibinfo {pages} {45} (\bibinfo {year}
  {2019})}\BibitemShut {NoStop}%
\bibitem [{\citenamefont {Fontenele}\ \emph {et~al.}(2019)\citenamefont
  {Fontenele}, \citenamefont {de~Vasconcelos}, \citenamefont {Feliciano},
  \citenamefont {Aguiar}, \citenamefont {Soares-Cunha}, \citenamefont
  {Coimbra}, \citenamefont {Dalla~Porta}, \citenamefont {Ribeiro},
  \citenamefont {Rodrigues}, \citenamefont {Sousa} \emph
  {et~al.}}]{Fontenele2019}%
  \BibitemOpen
  \bibfield  {author} {\bibinfo {author} {\bibfnamefont {A.~J.}\ \bibnamefont
  {Fontenele}}, \bibinfo {author} {\bibfnamefont {N.~A.}\ \bibnamefont
  {de~Vasconcelos}}, \bibinfo {author} {\bibfnamefont {T.}~\bibnamefont
  {Feliciano}}, \bibinfo {author} {\bibfnamefont {L.~A.}\ \bibnamefont
  {Aguiar}}, \bibinfo {author} {\bibfnamefont {C.}~\bibnamefont
  {Soares-Cunha}}, \bibinfo {author} {\bibfnamefont {B.}~\bibnamefont
  {Coimbra}}, \bibinfo {author} {\bibfnamefont {L.}~\bibnamefont
  {Dalla~Porta}}, \bibinfo {author} {\bibfnamefont {S.}~\bibnamefont
  {Ribeiro}}, \bibinfo {author} {\bibfnamefont {A.~J.}\ \bibnamefont
  {Rodrigues}}, \bibinfo {author} {\bibfnamefont {N.}~\bibnamefont {Sousa}},
  \emph {et~al.},\ }\href@noop {} {\bibfield  {journal} {\bibinfo  {journal}
  {Physical review letters}\ }\textbf {\bibinfo {volume} {122}},\ \bibinfo
  {pages} {208101} (\bibinfo {year} {2019})}\BibitemShut {NoStop}%
\bibitem [{\citenamefont {Miller}\ \emph {et~al.}(2019)\citenamefont {Miller},
  \citenamefont {Yu},\ and\ \citenamefont {Plenz}}]{Miller2019}%
  \BibitemOpen
  \bibfield  {author} {\bibinfo {author} {\bibfnamefont {S.~R.}\ \bibnamefont
  {Miller}}, \bibinfo {author} {\bibfnamefont {S.}~\bibnamefont {Yu}}, \ and\
  \bibinfo {author} {\bibfnamefont {D.}~\bibnamefont {Plenz}},\ }\href@noop {}
  {\bibfield  {journal} {\bibinfo  {journal} {Scientific reports}\ }\textbf
  {\bibinfo {volume} {9}},\ \bibinfo {pages} {1} (\bibinfo {year}
  {2019})}\BibitemShut {NoStop}%
\bibitem [{\citenamefont {Shaukat}\ and\ \citenamefont
  {Thivierge}(2016)}]{Shaukat2016}%
  \BibitemOpen
  \bibfield  {author} {\bibinfo {author} {\bibfnamefont {A.}~\bibnamefont
  {Shaukat}}\ and\ \bibinfo {author} {\bibfnamefont {J.-P.}\ \bibnamefont
  {Thivierge}},\ }\href@noop {} {\bibfield  {journal} {\bibinfo  {journal}
  {Frontiers in computational neuroscience}\ }\textbf {\bibinfo {volume}
  {10}},\ \bibinfo {pages} {29} (\bibinfo {year} {2016})}\BibitemShut {NoStop}%
\bibitem [{\citenamefont {Haldeman}\ and\ \citenamefont
  {Beggs}(2005)}]{Haldeman2005}%
  \BibitemOpen
  \bibfield  {author} {\bibinfo {author} {\bibfnamefont {C.}~\bibnamefont
  {Haldeman}}\ and\ \bibinfo {author} {\bibfnamefont {J.~M.}\ \bibnamefont
  {Beggs}},\ }\href@noop {} {\bibfield  {journal} {\bibinfo  {journal}
  {Physical review letters}\ }\textbf {\bibinfo {volume} {94}},\ \bibinfo
  {pages} {058101} (\bibinfo {year} {2005})}\BibitemShut {NoStop}%
\bibitem [{\citenamefont {Shew}\ \emph {et~al.}(2011)\citenamefont {Shew},
  \citenamefont {Yang}, \citenamefont {Yu}, \citenamefont {Roy},\ and\
  \citenamefont {Plenz}}]{Shew2011}%
  \BibitemOpen
  \bibfield  {author} {\bibinfo {author} {\bibfnamefont {W.~L.}\ \bibnamefont
  {Shew}}, \bibinfo {author} {\bibfnamefont {H.}~\bibnamefont {Yang}}, \bibinfo
  {author} {\bibfnamefont {S.}~\bibnamefont {Yu}}, \bibinfo {author}
  {\bibfnamefont {R.}~\bibnamefont {Roy}}, \ and\ \bibinfo {author}
  {\bibfnamefont {D.}~\bibnamefont {Plenz}},\ }\href@noop {} {\bibfield
  {journal} {\bibinfo  {journal} {Journal of neuroscience}\ }\textbf {\bibinfo
  {volume} {31}},\ \bibinfo {pages} {55} (\bibinfo {year} {2011})}\BibitemShut
  {NoStop}%
\bibitem [{\citenamefont {Li}\ \emph {et~al.}(2017)\citenamefont {Li},
  \citenamefont {Chen},\ and\ \citenamefont {Xue}}]{Li2017}%
  \BibitemOpen
  \bibfield  {author} {\bibinfo {author} {\bibfnamefont {X.}~\bibnamefont
  {Li}}, \bibinfo {author} {\bibfnamefont {Q.}~\bibnamefont {Chen}}, \ and\
  \bibinfo {author} {\bibfnamefont {F.}~\bibnamefont {Xue}},\ }\href@noop {}
  {\bibfield  {journal} {\bibinfo  {journal} {Philosophical Transactions of the
  Royal Society A: Mathematical, Physical and Engineering Sciences}\ }\textbf
  {\bibinfo {volume} {375}},\ \bibinfo {pages} {20160286} (\bibinfo {year}
  {2017})}\BibitemShut {NoStop}%
\bibitem [{\citenamefont {Williams-Garc{\'\i}a}\ \emph
  {et~al.}(2014)\citenamefont {Williams-Garc{\'\i}a}, \citenamefont {Moore},
  \citenamefont {Beggs},\ and\ \citenamefont {Ortiz}}]{Williams2014}%
  \BibitemOpen
  \bibfield  {author} {\bibinfo {author} {\bibfnamefont {R.~V.}\ \bibnamefont
  {Williams-Garc{\'\i}a}}, \bibinfo {author} {\bibfnamefont {M.}~\bibnamefont
  {Moore}}, \bibinfo {author} {\bibfnamefont {J.~M.}\ \bibnamefont {Beggs}}, \
  and\ \bibinfo {author} {\bibfnamefont {G.}~\bibnamefont {Ortiz}},\
  }\href@noop {} {\bibfield  {journal} {\bibinfo  {journal} {Physical Review
  E}\ }\textbf {\bibinfo {volume} {90}},\ \bibinfo {pages} {062714} (\bibinfo
  {year} {2014})}\BibitemShut {NoStop}%
\bibitem [{\citenamefont {Eurich}\ \emph {et~al.}(2002)\citenamefont {Eurich},
  \citenamefont {Herrmann},\ and\ \citenamefont {Ernst}}]{Eurich2002}%
  \BibitemOpen
  \bibfield  {author} {\bibinfo {author} {\bibfnamefont {C.~W.}\ \bibnamefont
  {Eurich}}, \bibinfo {author} {\bibfnamefont {J.~M.}\ \bibnamefont
  {Herrmann}}, \ and\ \bibinfo {author} {\bibfnamefont {U.~A.}\ \bibnamefont
  {Ernst}},\ }\href@noop {} {\bibfield  {journal} {\bibinfo  {journal}
  {Physical review E}\ }\textbf {\bibinfo {volume} {66}},\ \bibinfo {pages}
  {066137} (\bibinfo {year} {2002})}\BibitemShut {NoStop}%
\bibitem [{\citenamefont {Levina}\ \emph
  {et~al.}(2007{\natexlab{a}})\citenamefont {Levina}, \citenamefont {Ernst},\
  and\ \citenamefont {Herrmann}}]{Levina2007a}%
  \BibitemOpen
  \bibfield  {author} {\bibinfo {author} {\bibfnamefont {A.}~\bibnamefont
  {Levina}}, \bibinfo {author} {\bibfnamefont {U.}~\bibnamefont {Ernst}}, \
  and\ \bibinfo {author} {\bibfnamefont {J.~M.}\ \bibnamefont {Herrmann}},\
  }\href@noop {} {\bibfield  {journal} {\bibinfo  {journal} {Neurocomputing}\
  }\textbf {\bibinfo {volume} {70}},\ \bibinfo {pages} {1877} (\bibinfo {year}
  {2007}{\natexlab{a}})}\BibitemShut {NoStop}%
\bibitem [{\citenamefont {Wang}\ \emph {et~al.}(2011)\citenamefont {Wang},
  \citenamefont {Hilgetag},\ and\ \citenamefont {Zhou}}]{Wang2011}%
  \BibitemOpen
  \bibfield  {author} {\bibinfo {author} {\bibfnamefont {S.-J.}\ \bibnamefont
  {Wang}}, \bibinfo {author} {\bibfnamefont {C.}~\bibnamefont {Hilgetag}}, \
  and\ \bibinfo {author} {\bibfnamefont {C.}~\bibnamefont {Zhou}},\ }\href@noop
  {} {\bibfield  {journal} {\bibinfo  {journal} {Frontiers in computational
  neuroscience}\ }\textbf {\bibinfo {volume} {5}},\ \bibinfo {pages} {30}
  (\bibinfo {year} {2011})}\BibitemShut {NoStop}%
\bibitem [{\citenamefont {Poil}\ \emph {et~al.}(2012)\citenamefont {Poil},
  \citenamefont {Hardstone}, \citenamefont {Mansvelder},\ and\ \citenamefont
  {Linkenkaer-Hansen}}]{Poil2012}%
  \BibitemOpen
  \bibfield  {author} {\bibinfo {author} {\bibfnamefont {S.-S.}\ \bibnamefont
  {Poil}}, \bibinfo {author} {\bibfnamefont {R.}~\bibnamefont {Hardstone}},
  \bibinfo {author} {\bibfnamefont {H.~D.}\ \bibnamefont {Mansvelder}}, \ and\
  \bibinfo {author} {\bibfnamefont {K.}~\bibnamefont {Linkenkaer-Hansen}},\
  }\href@noop {} {\bibfield  {journal} {\bibinfo  {journal} {Journal of
  Neuroscience}\ }\textbf {\bibinfo {volume} {32}},\ \bibinfo {pages} {9817}
  (\bibinfo {year} {2012})}\BibitemShut {NoStop}%
\bibitem [{\citenamefont {Dalla~Porta}\ and\ \citenamefont
  {Copelli}(2019)}]{DallaPorta2019}%
  \BibitemOpen
  \bibfield  {author} {\bibinfo {author} {\bibfnamefont {L.}~\bibnamefont
  {Dalla~Porta}}\ and\ \bibinfo {author} {\bibfnamefont {M.}~\bibnamefont
  {Copelli}},\ }\href@noop {} {\bibfield  {journal} {\bibinfo  {journal} {PLoS
  computational biology}\ }\textbf {\bibinfo {volume} {15}},\ \bibinfo {pages}
  {e1006924} (\bibinfo {year} {2019})}\BibitemShut {NoStop}%
\bibitem [{\citenamefont {Haimovici}\ \emph {et~al.}(2013)\citenamefont
  {Haimovici}, \citenamefont {Tagliazucchi}, \citenamefont {Balenzuela},\ and\
  \citenamefont {Chialvo}}]{Haimovici2013}%
  \BibitemOpen
  \bibfield  {author} {\bibinfo {author} {\bibfnamefont {A.}~\bibnamefont
  {Haimovici}}, \bibinfo {author} {\bibfnamefont {E.}~\bibnamefont
  {Tagliazucchi}}, \bibinfo {author} {\bibfnamefont {P.}~\bibnamefont
  {Balenzuela}}, \ and\ \bibinfo {author} {\bibfnamefont {D.~R.}\ \bibnamefont
  {Chialvo}},\ }\href@noop {} {\bibfield  {journal} {\bibinfo  {journal}
  {Physical review letters}\ }\textbf {\bibinfo {volume} {110}},\ \bibinfo
  {pages} {178101} (\bibinfo {year} {2013})}\BibitemShut {NoStop}%
\bibitem [{\citenamefont {Rocha}\ \emph {et~al.}(2018)\citenamefont {Rocha},
  \citenamefont {Ko{\c{c}}illari}, \citenamefont {Suweis}, \citenamefont
  {Corbetta},\ and\ \citenamefont {Maritan}}]{Rocha2018}%
  \BibitemOpen
  \bibfield  {author} {\bibinfo {author} {\bibfnamefont {R.~P.}\ \bibnamefont
  {Rocha}}, \bibinfo {author} {\bibfnamefont {L.}~\bibnamefont
  {Ko{\c{c}}illari}}, \bibinfo {author} {\bibfnamefont {S.}~\bibnamefont
  {Suweis}}, \bibinfo {author} {\bibfnamefont {M.}~\bibnamefont {Corbetta}}, \
  and\ \bibinfo {author} {\bibfnamefont {A.}~\bibnamefont {Maritan}},\
  }\href@noop {} {\bibfield  {journal} {\bibinfo  {journal} {Scientific
  reports}\ }\textbf {\bibinfo {volume} {8}},\ \bibinfo {pages} {1} (\bibinfo
  {year} {2018})}\BibitemShut {NoStop}%
\bibitem [{\citenamefont {Benayoun}\ \emph {et~al.}(2010)\citenamefont
  {Benayoun}, \citenamefont {Cowan}, \citenamefont {van Drongelen},\ and\
  \citenamefont {Wallace}}]{Benayoun2010}%
  \BibitemOpen
  \bibfield  {author} {\bibinfo {author} {\bibfnamefont {M.}~\bibnamefont
  {Benayoun}}, \bibinfo {author} {\bibfnamefont {J.~D.}\ \bibnamefont {Cowan}},
  \bibinfo {author} {\bibfnamefont {W.}~\bibnamefont {van Drongelen}}, \ and\
  \bibinfo {author} {\bibfnamefont {E.}~\bibnamefont {Wallace}},\ }\href@noop
  {} {\bibfield  {journal} {\bibinfo  {journal} {PLoS computational biology}\
  }\textbf {\bibinfo {volume} {6}} (\bibinfo {year} {2010})}\BibitemShut
  {NoStop}%
\bibitem [{\citenamefont {Meisel}\ and\ \citenamefont
  {Gross}(2009)}]{Meisel2009}%
  \BibitemOpen
  \bibfield  {author} {\bibinfo {author} {\bibfnamefont {C.}~\bibnamefont
  {Meisel}}\ and\ \bibinfo {author} {\bibfnamefont {T.}~\bibnamefont {Gross}},\
  }\href@noop {} {\bibfield  {journal} {\bibinfo  {journal} {Physical Review
  E}\ }\textbf {\bibinfo {volume} {80}},\ \bibinfo {pages} {061917} (\bibinfo
  {year} {2009})}\BibitemShut {NoStop}%
\bibitem [{\citenamefont {Rubinov}\ \emph {et~al.}(2011)\citenamefont
  {Rubinov}, \citenamefont {Sporns}, \citenamefont {Thivierge},\ and\
  \citenamefont {Breakspear}}]{Rubinov2011}%
  \BibitemOpen
  \bibfield  {author} {\bibinfo {author} {\bibfnamefont {M.}~\bibnamefont
  {Rubinov}}, \bibinfo {author} {\bibfnamefont {O.}~\bibnamefont {Sporns}},
  \bibinfo {author} {\bibfnamefont {J.-P.}\ \bibnamefont {Thivierge}}, \ and\
  \bibinfo {author} {\bibfnamefont {M.}~\bibnamefont {Breakspear}},\
  }\href@noop {} {\bibfield  {journal} {\bibinfo  {journal} {PLoS computational
  biology}\ }\textbf {\bibinfo {volume} {7}} (\bibinfo {year}
  {2011})}\BibitemShut {NoStop}%
\bibitem [{\citenamefont {Teixeira}\ and\ \citenamefont
  {Shanahan}(2014)}]{Teixeira2014}%
  \BibitemOpen
  \bibfield  {author} {\bibinfo {author} {\bibfnamefont {F.~P.~P.}\
  \bibnamefont {Teixeira}}\ and\ \bibinfo {author} {\bibfnamefont
  {M.}~\bibnamefont {Shanahan}},\ }in\ \href@noop {} {\emph {\bibinfo
  {booktitle} {2014 International Joint Conference on Neural Networks
  (IJCNN)}}}\ (\bibinfo {organization} {IEEE},\ \bibinfo {year} {2014})\ pp.\
  \bibinfo {pages} {2383--2390}\BibitemShut {NoStop}%
\bibitem [{\citenamefont {Khoshkhou}\ and\ \citenamefont
  {Montakhab}(2019)}]{khoshkhou2019spike}%
  \BibitemOpen
  \bibfield  {author} {\bibinfo {author} {\bibfnamefont {M.}~\bibnamefont
  {Khoshkhou}}\ and\ \bibinfo {author} {\bibfnamefont {A.}~\bibnamefont
  {Montakhab}},\ }\href@noop {} {\bibfield  {journal} {\bibinfo  {journal}
  {Frontiers in systems neuroscience}\ }\textbf {\bibinfo {volume} {13}},\
  \bibinfo {pages} {73} (\bibinfo {year} {2019})}\BibitemShut {NoStop}%
\bibitem [{\citenamefont {Levina}\ \emph
  {et~al.}(2007{\natexlab{b}})\citenamefont {Levina}, \citenamefont
  {Herrmann},\ and\ \citenamefont {Geisel}}]{Levina2007b}%
  \BibitemOpen
  \bibfield  {author} {\bibinfo {author} {\bibfnamefont {A.}~\bibnamefont
  {Levina}}, \bibinfo {author} {\bibfnamefont {J.~M.}\ \bibnamefont
  {Herrmann}}, \ and\ \bibinfo {author} {\bibfnamefont {T.}~\bibnamefont
  {Geisel}},\ }\href@noop {} {\bibfield  {journal} {\bibinfo  {journal} {Nature
  physics}\ }\textbf {\bibinfo {volume} {3}},\ \bibinfo {pages} {857} (\bibinfo
  {year} {2007}{\natexlab{b}})}\BibitemShut {NoStop}%
\bibitem [{\citenamefont {Levina}\ \emph {et~al.}(2009)\citenamefont {Levina},
  \citenamefont {Herrmann},\ and\ \citenamefont {Geisel}}]{Levina2009}%
  \BibitemOpen
  \bibfield  {author} {\bibinfo {author} {\bibfnamefont {A.}~\bibnamefont
  {Levina}}, \bibinfo {author} {\bibfnamefont {J.~M.}\ \bibnamefont
  {Herrmann}}, \ and\ \bibinfo {author} {\bibfnamefont {T.}~\bibnamefont
  {Geisel}},\ }\href@noop {} {\bibfield  {journal} {\bibinfo  {journal}
  {Physical review letters}\ }\textbf {\bibinfo {volume} {102}},\ \bibinfo
  {pages} {118110} (\bibinfo {year} {2009})}\BibitemShut {NoStop}%
\bibitem [{\citenamefont {Millman}\ \emph {et~al.}(2010)\citenamefont
  {Millman}, \citenamefont {Mihalas}, \citenamefont {Kirkwood},\ and\
  \citenamefont {Niebur}}]{Millman2010}%
  \BibitemOpen
  \bibfield  {author} {\bibinfo {author} {\bibfnamefont {D.}~\bibnamefont
  {Millman}}, \bibinfo {author} {\bibfnamefont {S.}~\bibnamefont {Mihalas}},
  \bibinfo {author} {\bibfnamefont {A.}~\bibnamefont {Kirkwood}}, \ and\
  \bibinfo {author} {\bibfnamefont {E.}~\bibnamefont {Niebur}},\ }\href@noop {}
  {\bibfield  {journal} {\bibinfo  {journal} {Nature physics}\ }\textbf
  {\bibinfo {volume} {6}},\ \bibinfo {pages} {801} (\bibinfo {year}
  {2010})}\BibitemShut {NoStop}%
\bibitem [{\citenamefont {Shew}\ \emph {et~al.}(2015)\citenamefont {Shew},
  \citenamefont {Clawson}, \citenamefont {Pobst}, \citenamefont {Karimipanah},
  \citenamefont {Wright},\ and\ \citenamefont {Wessel}}]{Shew2015}%
  \BibitemOpen
  \bibfield  {author} {\bibinfo {author} {\bibfnamefont {W.~L.}\ \bibnamefont
  {Shew}}, \bibinfo {author} {\bibfnamefont {W.~P.}\ \bibnamefont {Clawson}},
  \bibinfo {author} {\bibfnamefont {J.}~\bibnamefont {Pobst}}, \bibinfo
  {author} {\bibfnamefont {Y.}~\bibnamefont {Karimipanah}}, \bibinfo {author}
  {\bibfnamefont {N.~C.}\ \bibnamefont {Wright}}, \ and\ \bibinfo {author}
  {\bibfnamefont {R.}~\bibnamefont {Wessel}},\ }\href@noop {} {\bibfield
  {journal} {\bibinfo  {journal} {Nature Physics}\ }\textbf {\bibinfo {volume}
  {11}},\ \bibinfo {pages} {659} (\bibinfo {year} {2015})}\BibitemShut
  {NoStop}%
\bibitem [{\citenamefont {Campos}\ \emph {et~al.}(2017)\citenamefont {Campos},
  \citenamefont {de~Andrade~Costa}, \citenamefont {Copelli},\ and\
  \citenamefont {Kinouchi}}]{Campos2017}%
  \BibitemOpen
  \bibfield  {author} {\bibinfo {author} {\bibfnamefont {J.~G.~F.}\
  \bibnamefont {Campos}}, \bibinfo {author} {\bibfnamefont {A.}~\bibnamefont
  {de~Andrade~Costa}}, \bibinfo {author} {\bibfnamefont {M.}~\bibnamefont
  {Copelli}}, \ and\ \bibinfo {author} {\bibfnamefont {O.}~\bibnamefont
  {Kinouchi}},\ }\href@noop {} {\bibfield  {journal} {\bibinfo  {journal}
  {Physical Review E}\ }\textbf {\bibinfo {volume} {95}},\ \bibinfo {pages}
  {042303} (\bibinfo {year} {2017})}\BibitemShut {NoStop}%
\bibitem [{\citenamefont {Zierenberg}\ \emph {et~al.}(2018)\citenamefont
  {Zierenberg}, \citenamefont {Wilting},\ and\ \citenamefont
  {Priesemann}}]{Zierenberg2018}%
  \BibitemOpen
  \bibfield  {author} {\bibinfo {author} {\bibfnamefont {J.}~\bibnamefont
  {Zierenberg}}, \bibinfo {author} {\bibfnamefont {J.}~\bibnamefont {Wilting}},
  \ and\ \bibinfo {author} {\bibfnamefont {V.}~\bibnamefont {Priesemann}},\
  }\href {\doibase 10.1103/PhysRevX.8.031018} {\bibfield  {journal} {\bibinfo
  {journal} {Phys. Rev. X}\ }\textbf {\bibinfo {volume} {8}},\ \bibinfo {pages}
  {031018} (\bibinfo {year} {2018})}\BibitemShut {NoStop}%
\bibitem [{\citenamefont {Magnasco}\ \emph {et~al.}(2009)\citenamefont
  {Magnasco}, \citenamefont {Piro},\ and\ \citenamefont
  {Cecchi}}]{Magnasco2009}%
  \BibitemOpen
  \bibfield  {author} {\bibinfo {author} {\bibfnamefont {M.~O.}\ \bibnamefont
  {Magnasco}}, \bibinfo {author} {\bibfnamefont {O.}~\bibnamefont {Piro}}, \
  and\ \bibinfo {author} {\bibfnamefont {G.~A.}\ \bibnamefont {Cecchi}},\
  }\href@noop {} {\bibfield  {journal} {\bibinfo  {journal} {Physical review
  letters}\ }\textbf {\bibinfo {volume} {102}},\ \bibinfo {pages} {258102}
  (\bibinfo {year} {2009})}\BibitemShut {NoStop}%
\bibitem [{\citenamefont {de~Arcangelis}\ \emph {et~al.}(2006)\citenamefont
  {de~Arcangelis}, \citenamefont {Perrone-Capano},\ and\ \citenamefont
  {Herrmann}}]{Arcangelis2006}%
  \BibitemOpen
  \bibfield  {author} {\bibinfo {author} {\bibfnamefont {L.}~\bibnamefont
  {de~Arcangelis}}, \bibinfo {author} {\bibfnamefont {C.}~\bibnamefont
  {Perrone-Capano}}, \ and\ \bibinfo {author} {\bibfnamefont {H.~J.}\
  \bibnamefont {Herrmann}},\ }\href@noop {} {\bibfield  {journal} {\bibinfo
  {journal} {Physical review letters}\ }\textbf {\bibinfo {volume} {96}},\
  \bibinfo {pages} {028107} (\bibinfo {year} {2006})}\BibitemShut {NoStop}%
\bibitem [{\citenamefont {Pellegrini}\ \emph {et~al.}(2007)\citenamefont
  {Pellegrini}, \citenamefont {de~Arcangelis}, \citenamefont {Herrmann},\ and\
  \citenamefont {Perrone-Capano}}]{pellegrini2007activity}%
  \BibitemOpen
  \bibfield  {author} {\bibinfo {author} {\bibfnamefont {G.~L.}\ \bibnamefont
  {Pellegrini}}, \bibinfo {author} {\bibfnamefont {L.}~\bibnamefont
  {de~Arcangelis}}, \bibinfo {author} {\bibfnamefont {H.~J.}\ \bibnamefont
  {Herrmann}}, \ and\ \bibinfo {author} {\bibfnamefont {C.}~\bibnamefont
  {Perrone-Capano}},\ }\href@noop {} {\bibfield  {journal} {\bibinfo  {journal}
  {Physical Review E}\ }\textbf {\bibinfo {volume} {76}},\ \bibinfo {pages}
  {016107} (\bibinfo {year} {2007})}\BibitemShut {NoStop}%
\bibitem [{\citenamefont {Abbott}\ and\ \citenamefont
  {Rohrkemper}(2007)}]{Abbott2007}%
  \BibitemOpen
  \bibfield  {author} {\bibinfo {author} {\bibfnamefont {L.}~\bibnamefont
  {Abbott}}\ and\ \bibinfo {author} {\bibfnamefont {R.}~\bibnamefont
  {Rohrkemper}},\ }\href@noop {} {\bibfield  {journal} {\bibinfo  {journal}
  {Progress in brain research}\ }\textbf {\bibinfo {volume} {165}},\ \bibinfo
  {pages} {13} (\bibinfo {year} {2007})}\BibitemShut {NoStop}%
\bibitem [{\citenamefont {Kossio}\ \emph {et~al.}(2018)\citenamefont {Kossio},
  \citenamefont {Goedeke}, \citenamefont {van~den Akker}, \citenamefont
  {Ibarz},\ and\ \citenamefont {Memmesheimer}}]{Kossio2018}%
  \BibitemOpen
  \bibfield  {author} {\bibinfo {author} {\bibfnamefont {F.~Y.~K.}\
  \bibnamefont {Kossio}}, \bibinfo {author} {\bibfnamefont {S.}~\bibnamefont
  {Goedeke}}, \bibinfo {author} {\bibfnamefont {B.}~\bibnamefont {van~den
  Akker}}, \bibinfo {author} {\bibfnamefont {B.}~\bibnamefont {Ibarz}}, \ and\
  \bibinfo {author} {\bibfnamefont {R.-M.}\ \bibnamefont {Memmesheimer}},\
  }\href@noop {} {\bibfield  {journal} {\bibinfo  {journal} {Physical review
  letters}\ }\textbf {\bibinfo {volume} {121}},\ \bibinfo {pages} {058301}
  (\bibinfo {year} {2018})}\BibitemShut {NoStop}%
\bibitem [{\citenamefont {Uhlig}\ \emph {et~al.}(2013)\citenamefont {Uhlig},
  \citenamefont {Levina}, \citenamefont {Geisel},\ and\ \citenamefont
  {Herrmann}}]{Uhlig2013}%
  \BibitemOpen
  \bibfield  {author} {\bibinfo {author} {\bibfnamefont {M.}~\bibnamefont
  {Uhlig}}, \bibinfo {author} {\bibfnamefont {A.}~\bibnamefont {Levina}},
  \bibinfo {author} {\bibfnamefont {T.}~\bibnamefont {Geisel}}, \ and\ \bibinfo
  {author} {\bibfnamefont {M.}~\bibnamefont {Herrmann}},\ }\href@noop {}
  {\bibfield  {journal} {\bibinfo  {journal} {Frontiers in computational
  neuroscience}\ }\textbf {\bibinfo {volume} {7}},\ \bibinfo {pages} {87}
  (\bibinfo {year} {2013})}\BibitemShut {NoStop}%
\bibitem [{\citenamefont {Stepp}\ \emph {et~al.}(2015)\citenamefont {Stepp},
  \citenamefont {Plenz},\ and\ \citenamefont {Srinivasa}}]{Stepp2015}%
  \BibitemOpen
  \bibfield  {author} {\bibinfo {author} {\bibfnamefont {N.}~\bibnamefont
  {Stepp}}, \bibinfo {author} {\bibfnamefont {D.}~\bibnamefont {Plenz}}, \ and\
  \bibinfo {author} {\bibfnamefont {N.}~\bibnamefont {Srinivasa}},\ }\href@noop
  {} {\bibfield  {journal} {\bibinfo  {journal} {PLoS computational biology}\
  }\textbf {\bibinfo {volume} {11}} (\bibinfo {year} {2015})}\BibitemShut
  {NoStop}%
\bibitem [{\citenamefont {Brochini}\ \emph {et~al.}(2016)\citenamefont
  {Brochini}, \citenamefont {de~Andrade~Costa}, \citenamefont {Abadi},
  \citenamefont {Roque}, \citenamefont {Stolfi},\ and\ \citenamefont
  {Kinouchi}}]{brochini2016phase}%
  \BibitemOpen
  \bibfield  {author} {\bibinfo {author} {\bibfnamefont {L.}~\bibnamefont
  {Brochini}}, \bibinfo {author} {\bibfnamefont {A.}~\bibnamefont
  {de~Andrade~Costa}}, \bibinfo {author} {\bibfnamefont {M.}~\bibnamefont
  {Abadi}}, \bibinfo {author} {\bibfnamefont {A.~C.}\ \bibnamefont {Roque}},
  \bibinfo {author} {\bibfnamefont {J.}~\bibnamefont {Stolfi}}, \ and\ \bibinfo
  {author} {\bibfnamefont {O.}~\bibnamefont {Kinouchi}},\ }\href@noop {}
  {\bibfield  {journal} {\bibinfo  {journal} {Scientific reports}\ }\textbf
  {\bibinfo {volume} {6}},\ \bibinfo {pages} {1} (\bibinfo {year}
  {2016})}\BibitemShut {NoStop}%
\bibitem [{\citenamefont {Costa}\ \emph {et~al.}(2017)\citenamefont {Costa},
  \citenamefont {Brochini},\ and\ \citenamefont {Kinouchi}}]{costa2017self}%
  \BibitemOpen
  \bibfield  {author} {\bibinfo {author} {\bibfnamefont {A.~A.}\ \bibnamefont
  {Costa}}, \bibinfo {author} {\bibfnamefont {L.}~\bibnamefont {Brochini}}, \
  and\ \bibinfo {author} {\bibfnamefont {O.}~\bibnamefont {Kinouchi}},\
  }\href@noop {} {\bibfield  {journal} {\bibinfo  {journal} {Entropy}\ }\textbf
  {\bibinfo {volume} {19}},\ \bibinfo {pages} {399} (\bibinfo {year}
  {2017})}\BibitemShut {NoStop}%
\bibitem [{\citenamefont {Hernandez-Urbina}\ and\ \citenamefont
  {Herrmann}(2017)}]{hernandez2017self}%
  \BibitemOpen
  \bibfield  {author} {\bibinfo {author} {\bibfnamefont {V.}~\bibnamefont
  {Hernandez-Urbina}}\ and\ \bibinfo {author} {\bibfnamefont {J.~M.}\
  \bibnamefont {Herrmann}},\ }\href@noop {} {\bibfield  {journal} {\bibinfo
  {journal} {Frontiers in Physics}\ }\textbf {\bibinfo {volume} {4}},\ \bibinfo
  {pages} {54} (\bibinfo {year} {2017})}\BibitemShut {NoStop}%
\bibitem [{\citenamefont {Girardi-Schappo}\ \emph {et~al.}(2020)\citenamefont
  {Girardi-Schappo}, \citenamefont {Brochini}, \citenamefont {Costa},
  \citenamefont {Carvalho},\ and\ \citenamefont
  {Kinouchi}}]{GirardiSchappo2020}%
  \BibitemOpen
  \bibfield  {author} {\bibinfo {author} {\bibfnamefont {M.}~\bibnamefont
  {Girardi-Schappo}}, \bibinfo {author} {\bibfnamefont {L.}~\bibnamefont
  {Brochini}}, \bibinfo {author} {\bibfnamefont {A.~A.}\ \bibnamefont {Costa}},
  \bibinfo {author} {\bibfnamefont {T.~T.}\ \bibnamefont {Carvalho}}, \ and\
  \bibinfo {author} {\bibfnamefont {O.}~\bibnamefont {Kinouchi}},\ }\href@noop
  {} {\bibfield  {journal} {\bibinfo  {journal} {Physical Review Research}\
  }\textbf {\bibinfo {volume} {2}},\ \bibinfo {pages} {012042} (\bibinfo {year}
  {2020})}\BibitemShut {NoStop}%
\bibitem [{\citenamefont {Zeraati}\ \emph {et~al.}(2021)\citenamefont
  {Zeraati}, \citenamefont {Priesemann},\ and\ \citenamefont
  {Levina}}]{zeraati2021}%
  \BibitemOpen
  \bibfield  {author} {\bibinfo {author} {\bibfnamefont {R.}~\bibnamefont
  {Zeraati}}, \bibinfo {author} {\bibfnamefont {V.}~\bibnamefont {Priesemann}},
  \ and\ \bibinfo {author} {\bibfnamefont {A.}~\bibnamefont {Levina}},\ }\href
  {\doibase 10.3389/fphy.2021.619661} {\bibfield  {journal} {\bibinfo
  {journal} {Frontiers in Physics}\ }\textbf {\bibinfo {volume} {9}} (\bibinfo
  {year} {2021}),\ 10.3389/fphy.2021.619661}\BibitemShut {NoStop}%
\bibitem [{\citenamefont {Scarpetta}\ \emph {et~al.}(2018)\citenamefont
  {Scarpetta}, \citenamefont {Apicella}, \citenamefont {Minati},\ and\
  \citenamefont {de~Candia}}]{Scarpetta2018}%
  \BibitemOpen
  \bibfield  {author} {\bibinfo {author} {\bibfnamefont {S.}~\bibnamefont
  {Scarpetta}}, \bibinfo {author} {\bibfnamefont {I.}~\bibnamefont {Apicella}},
  \bibinfo {author} {\bibfnamefont {L.}~\bibnamefont {Minati}}, \ and\ \bibinfo
  {author} {\bibfnamefont {A.}~\bibnamefont {de~Candia}},\ }\href@noop {}
  {\bibfield  {journal} {\bibinfo  {journal} {Physical Review E}\ }\textbf
  {\bibinfo {volume} {97}},\ \bibinfo {pages} {062305} (\bibinfo {year}
  {2018})}\BibitemShut {NoStop}%
\bibitem [{\citenamefont {van Kessenich}\ \emph {et~al.}(2018)\citenamefont
  {van Kessenich}, \citenamefont {Lukovi{\'c}}, \citenamefont {De~Arcangelis},\
  and\ \citenamefont {Herrmann}}]{Kessenich2018}%
  \BibitemOpen
  \bibfield  {author} {\bibinfo {author} {\bibfnamefont {L.~M.}\ \bibnamefont
  {van Kessenich}}, \bibinfo {author} {\bibfnamefont {M.}~\bibnamefont
  {Lukovi{\'c}}}, \bibinfo {author} {\bibfnamefont {L.}~\bibnamefont
  {De~Arcangelis}}, \ and\ \bibinfo {author} {\bibfnamefont {H.~J.}\
  \bibnamefont {Herrmann}},\ }\href@noop {} {\bibfield  {journal} {\bibinfo
  {journal} {Physical Review E}\ }\textbf {\bibinfo {volume} {97}},\ \bibinfo
  {pages} {032312} (\bibinfo {year} {2018})}\BibitemShut {NoStop}%
\bibitem [{\citenamefont {Del~Papa}\ \emph {et~al.}(2017)\citenamefont
  {Del~Papa}, \citenamefont {Priesemann},\ and\ \citenamefont
  {Triesch}}]{DelPapa2017}%
  \BibitemOpen
  \bibfield  {author} {\bibinfo {author} {\bibfnamefont {B.}~\bibnamefont
  {Del~Papa}}, \bibinfo {author} {\bibfnamefont {V.}~\bibnamefont
  {Priesemann}}, \ and\ \bibinfo {author} {\bibfnamefont {J.}~\bibnamefont
  {Triesch}},\ }\href@noop {} {\bibfield  {journal} {\bibinfo  {journal} {PloS
  one}\ }\textbf {\bibinfo {volume} {12}} (\bibinfo {year} {2017})}\BibitemShut
  {NoStop}%
\bibitem [{\citenamefont {Shadlen}\ and\ \citenamefont
  {Newsome}(1994)}]{Shadlen1994}%
  \BibitemOpen
  \bibfield  {author} {\bibinfo {author} {\bibfnamefont {M.~N.}\ \bibnamefont
  {Shadlen}}\ and\ \bibinfo {author} {\bibfnamefont {W.~T.}\ \bibnamefont
  {Newsome}},\ }\href@noop {} {\bibfield  {journal} {\bibinfo  {journal}
  {Current opinion in neurobiology}\ }\textbf {\bibinfo {volume} {4}},\
  \bibinfo {pages} {569} (\bibinfo {year} {1994})}\BibitemShut {NoStop}%
\bibitem [{\citenamefont {Van~Vreeswijk}\ and\ \citenamefont
  {Sompolinsky}(1996)}]{Vreeswijk1996}%
  \BibitemOpen
  \bibfield  {author} {\bibinfo {author} {\bibfnamefont {C.}~\bibnamefont
  {Van~Vreeswijk}}\ and\ \bibinfo {author} {\bibfnamefont {H.}~\bibnamefont
  {Sompolinsky}},\ }\href@noop {} {\bibfield  {journal} {\bibinfo  {journal}
  {Science}\ }\textbf {\bibinfo {volume} {274}},\ \bibinfo {pages} {1724}
  (\bibinfo {year} {1996})}\BibitemShut {NoStop}%
\bibitem [{\citenamefont {Amit}\ and\ \citenamefont {Brunel}(1997)}]{Amit1997}%
  \BibitemOpen
  \bibfield  {author} {\bibinfo {author} {\bibfnamefont {D.~J.}\ \bibnamefont
  {Amit}}\ and\ \bibinfo {author} {\bibfnamefont {N.}~\bibnamefont {Brunel}},\
  }\href@noop {} {\bibfield  {journal} {\bibinfo  {journal} {Cerebral cortex
  (New York, NY: 1991)}\ }\textbf {\bibinfo {volume} {7}},\ \bibinfo {pages}
  {237} (\bibinfo {year} {1997})}\BibitemShut {NoStop}%
\bibitem [{\citenamefont {Shadlen}\ and\ \citenamefont
  {Newsome}(1998)}]{Shadlen1998}%
  \BibitemOpen
  \bibfield  {author} {\bibinfo {author} {\bibfnamefont {M.~N.}\ \bibnamefont
  {Shadlen}}\ and\ \bibinfo {author} {\bibfnamefont {W.~T.}\ \bibnamefont
  {Newsome}},\ }\href@noop {} {\bibfield  {journal} {\bibinfo  {journal}
  {Journal of neuroscience}\ }\textbf {\bibinfo {volume} {18}},\ \bibinfo
  {pages} {3870} (\bibinfo {year} {1998})}\BibitemShut {NoStop}%
\bibitem [{\citenamefont {Brunel}(2000)}]{Brunel2000}%
  \BibitemOpen
  \bibfield  {author} {\bibinfo {author} {\bibfnamefont {N.}~\bibnamefont
  {Brunel}},\ }\href@noop {} {\bibfield  {journal} {\bibinfo  {journal}
  {Journal of computational neuroscience}\ }\textbf {\bibinfo {volume} {8}},\
  \bibinfo {pages} {183} (\bibinfo {year} {2000})}\BibitemShut {NoStop}%
\bibitem [{\citenamefont {Shu}\ \emph {et~al.}(2003)\citenamefont {Shu},
  \citenamefont {Hasenstaub},\ and\ \citenamefont {McCormick}}]{Shu2003}%
  \BibitemOpen
  \bibfield  {author} {\bibinfo {author} {\bibfnamefont {Y.}~\bibnamefont
  {Shu}}, \bibinfo {author} {\bibfnamefont {A.}~\bibnamefont {Hasenstaub}}, \
  and\ \bibinfo {author} {\bibfnamefont {D.~A.}\ \bibnamefont {McCormick}},\
  }\href@noop {} {\bibfield  {journal} {\bibinfo  {journal} {Nature}\ }\textbf
  {\bibinfo {volume} {423}},\ \bibinfo {pages} {288} (\bibinfo {year}
  {2003})}\BibitemShut {NoStop}%
\bibitem [{\citenamefont {Wehr}\ and\ \citenamefont {Zador}(2003)}]{Wehr2003}%
  \BibitemOpen
  \bibfield  {author} {\bibinfo {author} {\bibfnamefont {M.}~\bibnamefont
  {Wehr}}\ and\ \bibinfo {author} {\bibfnamefont {A.~M.}\ \bibnamefont
  {Zador}},\ }\href@noop {} {\bibfield  {journal} {\bibinfo  {journal}
  {Nature}\ }\textbf {\bibinfo {volume} {426}},\ \bibinfo {pages} {442}
  (\bibinfo {year} {2003})}\BibitemShut {NoStop}%
\bibitem [{\citenamefont {Isaacson}\ and\ \citenamefont
  {Scanziani}(2011)}]{Isaacson2011}%
  \BibitemOpen
  \bibfield  {author} {\bibinfo {author} {\bibfnamefont {J.~S.}\ \bibnamefont
  {Isaacson}}\ and\ \bibinfo {author} {\bibfnamefont {M.}~\bibnamefont
  {Scanziani}},\ }\href@noop {} {\bibfield  {journal} {\bibinfo  {journal}
  {Neuron}\ }\textbf {\bibinfo {volume} {72}},\ \bibinfo {pages} {231}
  (\bibinfo {year} {2011})}\BibitemShut {NoStop}%
\bibitem [{\citenamefont {Den{\`e}ve}\ and\ \citenamefont
  {Machens}(2016)}]{Deneve2016}%
  \BibitemOpen
  \bibfield  {author} {\bibinfo {author} {\bibfnamefont {S.}~\bibnamefont
  {Den{\`e}ve}}\ and\ \bibinfo {author} {\bibfnamefont {C.~K.}\ \bibnamefont
  {Machens}},\ }\href@noop {} {\bibfield  {journal} {\bibinfo  {journal}
  {Nature neuroscience}\ }\textbf {\bibinfo {volume} {19}},\ \bibinfo {pages}
  {375} (\bibinfo {year} {2016})}\BibitemShut {NoStop}%
\bibitem [{\citenamefont {Sahara}\ \emph {et~al.}(2012)\citenamefont {Sahara},
  \citenamefont {Yanagawa}, \citenamefont {O{\textquoteright}Leary},\ and\
  \citenamefont {Stevens}}]{Sahara2012}%
  \BibitemOpen
  \bibfield  {author} {\bibinfo {author} {\bibfnamefont {S.}~\bibnamefont
  {Sahara}}, \bibinfo {author} {\bibfnamefont {Y.}~\bibnamefont {Yanagawa}},
  \bibinfo {author} {\bibfnamefont {D.~D.~M.}\ \bibnamefont
  {O{\textquoteright}Leary}}, \ and\ \bibinfo {author} {\bibfnamefont {C.~F.}\
  \bibnamefont {Stevens}},\ }\href {\doibase 10.1523/JNEUROSCI.6412-11.2012}
  {\bibfield  {journal} {\bibinfo  {journal} {Journal of Neuroscience}\
  }\textbf {\bibinfo {volume} {32}},\ \bibinfo {pages} {4755} (\bibinfo {year}
  {2012})},\ \Eprint
  {http://arxiv.org/abs/https://www.jneurosci.org/content/32/14/4755.full.pdf}
  {https://www.jneurosci.org/content/32/14/4755.full.pdf} \BibitemShut
  {NoStop}%
\bibitem [{\citenamefont {Bornholdt}\ and\ \citenamefont
  {Rohlf}(2000)}]{bornholdt2000topological}%
  \BibitemOpen
  \bibfield  {author} {\bibinfo {author} {\bibfnamefont {S.}~\bibnamefont
  {Bornholdt}}\ and\ \bibinfo {author} {\bibfnamefont {T.}~\bibnamefont
  {Rohlf}},\ }\href@noop {} {\bibfield  {journal} {\bibinfo  {journal}
  {Physical Review Letters}\ }\textbf {\bibinfo {volume} {84}},\ \bibinfo
  {pages} {6114} (\bibinfo {year} {2000})}\BibitemShut {NoStop}%
\bibitem [{\citenamefont {Bornholdt}\ and\ \citenamefont
  {R{\"o}hl}(2003)}]{bornholdt2003self}%
  \BibitemOpen
  \bibfield  {author} {\bibinfo {author} {\bibfnamefont {S.}~\bibnamefont
  {Bornholdt}}\ and\ \bibinfo {author} {\bibfnamefont {T.}~\bibnamefont
  {R{\"o}hl}},\ }\href@noop {} {\bibfield  {journal} {\bibinfo  {journal}
  {Physical Review E}\ }\textbf {\bibinfo {volume} {67}},\ \bibinfo {pages}
  {066118} (\bibinfo {year} {2003})}\BibitemShut {NoStop}%
\bibitem [{\citenamefont {Gross}\ and\ \citenamefont
  {Blasius}(2008)}]{gross2008adaptive}%
  \BibitemOpen
  \bibfield  {author} {\bibinfo {author} {\bibfnamefont {T.}~\bibnamefont
  {Gross}}\ and\ \bibinfo {author} {\bibfnamefont {B.}~\bibnamefont
  {Blasius}},\ }\href@noop {} {\bibfield  {journal} {\bibinfo  {journal}
  {Journal of the Royal Society Interface}\ }\textbf {\bibinfo {volume} {5}},\
  \bibinfo {pages} {259} (\bibinfo {year} {2008})}\BibitemShut {NoStop}%
\bibitem [{\citenamefont {Gross}\ and\ \citenamefont
  {Sayama}(2009)}]{gross2009adaptive}%
  \BibitemOpen
  \bibfield  {author} {\bibinfo {author} {\bibfnamefont {T.}~\bibnamefont
  {Gross}}\ and\ \bibinfo {author} {\bibfnamefont {H.}~\bibnamefont {Sayama}},\
  }\href@noop {} {\emph {\bibinfo {title} {Adaptive Networks: Theory, Models
  and Applications}}}\ (\bibinfo  {publisher} {Springer Science \& Business
  Media},\ \bibinfo {year} {2009})\BibitemShut {NoStop}%
\bibitem [{\citenamefont {Droste}\ \emph {et~al.}(2013)\citenamefont {Droste},
  \citenamefont {Do},\ and\ \citenamefont {Gross}}]{droste2013}%
  \BibitemOpen
  \bibfield  {author} {\bibinfo {author} {\bibfnamefont {F.}~\bibnamefont
  {Droste}}, \bibinfo {author} {\bibfnamefont {A.-L.}\ \bibnamefont {Do}}, \
  and\ \bibinfo {author} {\bibfnamefont {T.}~\bibnamefont {Gross}},\
  }\href@noop {} {\bibfield  {journal} {\bibinfo  {journal} {Journal of The
  Royal Society Interface}\ }\textbf {\bibinfo {volume} {10}},\ \bibinfo
  {pages} {20120558} (\bibinfo {year} {2013})}\BibitemShut {NoStop}%
\bibitem [{\citenamefont {K{\"u}rten}(1988{\natexlab{b}})}]{Kuerten1988}%
  \BibitemOpen
  \bibfield  {author} {\bibinfo {author} {\bibfnamefont {K.~E.}\ \bibnamefont
  {K{\"u}rten}},\ }\href@noop {} {\bibfield  {journal} {\bibinfo  {journal}
  {Physics Letters A}\ }\textbf {\bibinfo {volume} {129}},\ \bibinfo {pages}
  {157} (\bibinfo {year} {1988}{\natexlab{b}})}\BibitemShut {NoStop}%
\bibitem [{\citenamefont {Neto}\ \emph {et~al.}(2017)\citenamefont {Neto},
  \citenamefont {de~Aguiar}, \citenamefont {Brum},\ and\ \citenamefont
  {Bornholdt}}]{Neto2017}%
  \BibitemOpen
  \bibfield  {author} {\bibinfo {author} {\bibfnamefont {J.~P.}\ \bibnamefont
  {Neto}}, \bibinfo {author} {\bibfnamefont {M.~A.}\ \bibnamefont {de~Aguiar}},
  \bibinfo {author} {\bibfnamefont {J.~A.}\ \bibnamefont {Brum}}, \ and\
  \bibinfo {author} {\bibfnamefont {S.}~\bibnamefont {Bornholdt}},\ }\href@noop
  {} {\bibfield  {journal} {\bibinfo  {journal} {arXiv preprint
  arXiv:1712.08816}\ } (\bibinfo {year} {2017})}\BibitemShut {NoStop}%
\bibitem [{\citenamefont {Baumgarten}\ and\ \citenamefont
  {Bornholdt}(2019{\natexlab{a}})}]{Baumgarten2019}%
  \BibitemOpen
  \bibfield  {author} {\bibinfo {author} {\bibfnamefont {L.}~\bibnamefont
  {Baumgarten}}\ and\ \bibinfo {author} {\bibfnamefont {S.}~\bibnamefont
  {Bornholdt}},\ }\href@noop {} {\bibfield  {journal} {\bibinfo  {journal}
  {Physical Review E}\ }\textbf {\bibinfo {volume} {100}},\ \bibinfo {pages}
  {010301} (\bibinfo {year} {2019}{\natexlab{a}})}\BibitemShut {NoStop}%
\bibitem [{\citenamefont {Jiang}\ \emph {et~al.}(2005)\citenamefont {Jiang},
  \citenamefont {Huang}, \citenamefont {Morales},\ and\ \citenamefont
  {Kirkwood}}]{Jiang2005}%
  \BibitemOpen
  \bibfield  {author} {\bibinfo {author} {\bibfnamefont {B.}~\bibnamefont
  {Jiang}}, \bibinfo {author} {\bibfnamefont {Z.~J.}\ \bibnamefont {Huang}},
  \bibinfo {author} {\bibfnamefont {B.}~\bibnamefont {Morales}}, \ and\
  \bibinfo {author} {\bibfnamefont {A.}~\bibnamefont {Kirkwood}},\ }\href
  {\doibase https://doi.org/10.1016/j.brainresrev.2005.05.007} {\bibfield
  {journal} {\bibinfo  {journal} {Brain Research Reviews}\ }\textbf {\bibinfo
  {volume} {50}},\ \bibinfo {pages} {126 } (\bibinfo {year}
  {2005})}\BibitemShut {NoStop}%
\bibitem [{\citenamefont {Landmann}\ \emph {et~al.}(2021)\citenamefont
  {Landmann}, \citenamefont {Baumgarten},\ and\ \citenamefont
  {Bornholdt}}]{landmann2021}%
  \BibitemOpen
  \bibfield  {author} {\bibinfo {author} {\bibfnamefont {S.}~\bibnamefont
  {Landmann}}, \bibinfo {author} {\bibfnamefont {L.}~\bibnamefont
  {Baumgarten}}, \ and\ \bibinfo {author} {\bibfnamefont {S.}~\bibnamefont
  {Bornholdt}},\ }\href@noop {} {\bibfield  {journal} {\bibinfo  {journal}
  {Physical Review E}\ }\textbf {\bibinfo {volume} {103}},\ \bibinfo {pages}
  {032304} (\bibinfo {year} {2021})}\BibitemShut {NoStop}%
\bibitem [{\citenamefont {Baumgarten}\ and\ \citenamefont
  {Bornholdt}(2019{\natexlab{b}})}]{BaumgartenBornholdt2019}%
  \BibitemOpen
  \bibfield  {author} {\bibinfo {author} {\bibfnamefont {L.}~\bibnamefont
  {Baumgarten}}\ and\ \bibinfo {author} {\bibfnamefont {S.}~\bibnamefont
  {Bornholdt}},\ }\href {\doibase 10.1103/PhysRevE.100.010301} {\bibfield
  {journal} {\bibinfo  {journal} {Phys. Rev. E}\ }\textbf {\bibinfo {volume}
  {100}},\ \bibinfo {pages} {010301} (\bibinfo {year}
  {2019}{\natexlab{b}})}\BibitemShut {NoStop}%
\bibitem [{\citenamefont {Luque}\ and\ \citenamefont
  {Sol\'e}(1997)}]{LuqueSole1997}%
  \BibitemOpen
  \bibfield  {author} {\bibinfo {author} {\bibfnamefont {B.}~\bibnamefont
  {Luque}}\ and\ \bibinfo {author} {\bibfnamefont {R.}~\bibnamefont {Sol\'e}},\
  }\href@noop {} {\bibfield  {journal} {\bibinfo  {journal} {Phys. Rev. E}\
  }\textbf {\bibinfo {volume} {55}},\ \bibinfo {pages} {257} (\bibinfo {year}
  {1997})}\BibitemShut {NoStop}%
\bibitem [{\citenamefont {Shmulevich}\ and\ \citenamefont
  {Kauffman}(2004)}]{ShmulevichKauffman2004}%
  \BibitemOpen
  \bibfield  {author} {\bibinfo {author} {\bibfnamefont {I.}~\bibnamefont
  {Shmulevich}}\ and\ \bibinfo {author} {\bibfnamefont {S.}~\bibnamefont
  {Kauffman}},\ }\href@noop {} {\bibfield  {journal} {\bibinfo  {journal}
  {Phys. Rev. Lett.}\ }\textbf {\bibinfo {volume} {93}},\ \bibinfo {pages}
  {048701} (\bibinfo {year} {2004})}\BibitemShut {NoStop}%
\bibitem [{\citenamefont {Sethna}\ \emph {et~al.}(2001)\citenamefont {Sethna},
  \citenamefont {Dahmen},\ and\ \citenamefont {Myers}}]{sethna2001}%
  \BibitemOpen
  \bibfield  {author} {\bibinfo {author} {\bibfnamefont {J.~P.}\ \bibnamefont
  {Sethna}}, \bibinfo {author} {\bibfnamefont {K.~A.}\ \bibnamefont {Dahmen}},
  \ and\ \bibinfo {author} {\bibfnamefont {C.~R.}\ \bibnamefont {Myers}},\
  }\href@noop {} {\bibfield  {journal} {\bibinfo  {journal} {Nature}\ }\textbf
  {\bibinfo {volume} {410}},\ \bibinfo {pages} {242} (\bibinfo {year}
  {2001})}\BibitemShut {NoStop}%
\bibitem [{\citenamefont {Clauset}\ \emph {et~al.}(2009)\citenamefont
  {Clauset}, \citenamefont {Shalizi},\ and\ \citenamefont
  {Newman}}]{Clauset2009}%
  \BibitemOpen
  \bibfield  {author} {\bibinfo {author} {\bibfnamefont {A.}~\bibnamefont
  {Clauset}}, \bibinfo {author} {\bibfnamefont {C.~R.}\ \bibnamefont
  {Shalizi}}, \ and\ \bibinfo {author} {\bibfnamefont {M.~E.}\ \bibnamefont
  {Newman}},\ }\href@noop {} {\bibfield  {journal} {\bibinfo  {journal} {SIAM
  Review}\ }\textbf {\bibinfo {volume} {51}},\ \bibinfo {pages} {661} (\bibinfo
  {year} {2009})}\BibitemShut {NoStop}%
\bibitem [{\citenamefont {Lin}\ and\ \citenamefont {Chen}(2005)}]{Lin2005}%
  \BibitemOpen
  \bibfield  {author} {\bibinfo {author} {\bibfnamefont {M.}~\bibnamefont
  {Lin}}\ and\ \bibinfo {author} {\bibfnamefont {T.}~\bibnamefont {Chen}},\
  }\href@noop {} {\bibfield  {journal} {\bibinfo  {journal} {Physical Review
  E}\ }\textbf {\bibinfo {volume} {71}},\ \bibinfo {pages} {016133} (\bibinfo
  {year} {2005})}\BibitemShut {NoStop}%
\bibitem [{\citenamefont {Pazzini}\ \emph {et~al.}(2021)\citenamefont
  {Pazzini}, \citenamefont {Kinouchi},\ and\ \citenamefont
  {Costa}}]{Pazzini2021}%
  \BibitemOpen
  \bibfield  {author} {\bibinfo {author} {\bibfnamefont {R.}~\bibnamefont
  {Pazzini}}, \bibinfo {author} {\bibfnamefont {O.}~\bibnamefont {Kinouchi}}, \
  and\ \bibinfo {author} {\bibfnamefont {A.~A.}\ \bibnamefont {Costa}},\ }\href
  {\doibase 10.1103/PhysRevE.104.014137} {\bibfield  {journal} {\bibinfo
  {journal} {Phys. Rev. E}\ }\textbf {\bibinfo {volume} {104}},\ \bibinfo
  {pages} {014137} (\bibinfo {year} {2021})}\BibitemShut {NoStop}%
\bibitem [{\citenamefont {Sukenik}\ \emph {et~al.}(2021)\citenamefont
  {Sukenik}, \citenamefont {Vinogradov}, \citenamefont {Weinreb}, \citenamefont
  {Segal}, \citenamefont {Levina},\ and\ \citenamefont {Moses}}]{sukenik2021}%
  \BibitemOpen
  \bibfield  {author} {\bibinfo {author} {\bibfnamefont {N.}~\bibnamefont
  {Sukenik}}, \bibinfo {author} {\bibfnamefont {O.}~\bibnamefont {Vinogradov}},
  \bibinfo {author} {\bibfnamefont {E.}~\bibnamefont {Weinreb}}, \bibinfo
  {author} {\bibfnamefont {M.}~\bibnamefont {Segal}}, \bibinfo {author}
  {\bibfnamefont {A.}~\bibnamefont {Levina}}, \ and\ \bibinfo {author}
  {\bibfnamefont {E.}~\bibnamefont {Moses}},\ }\href@noop {} {\bibfield
  {journal} {\bibinfo  {journal} {Proceedings of the National Academy of
  Sciences}\ }\textbf {\bibinfo {volume} {118}} (\bibinfo {year}
  {2021})}\BibitemShut {NoStop}%
\bibitem [{\citenamefont {Gross}(2021)}]{gross2021not}%
  \BibitemOpen
  \bibfield  {author} {\bibinfo {author} {\bibfnamefont {T.}~\bibnamefont
  {Gross}},\ }\href@noop {} {\bibfield  {journal} {\bibinfo  {journal}
  {Frontiers in Neural Circuits}\ }\textbf {\bibinfo {volume} {15}},\ \bibinfo
  {pages} {7} (\bibinfo {year} {2021})}\BibitemShut {NoStop}%
\end{thebibliography}%
\end{document}